%% file: main.tex
\documentclass[12pt,onecolumn,draftcls]{IEEEtran}
\input{packages.tex}

\begin{document}
%
\title{Blocking time under basic priority inheritance: Polynomial bound and exact computation}
%
%
%

\author{Paolo~Torroni, \and
Zeynep~Kiziltan, \and
Eugenio~Faldella
\thanks{Department
of Computer Science and Engineering, University of Bologna, Italy e-mail: name.surname@unibo.it.}
}

\maketitle

\begin{abstract}
The Priority Inheritance Protocol (PIP) is arguably the best-known protocol for resource sharing under real-time constraints. Its importance in modern applications is undisputed. Nevertheless, because jobs may be blocked under PIP for a variety of reasons, determining a job's maximum blocking time could be difficult, and thus far no exact method has been proposed that does it. Existing analysis methods are inefficient, inaccurate, and of limited applicability. This article proposes a new characterization of the problem, thus allowing a polynomial method for bounding the blocking time, and an exact, optimally efficient method for blocking time computation under priority inheritance that have a general applicability.
\end{abstract}


%
\IEEEpeerreviewmaketitle

\section{Introduction}
\label{sec:intro}
\input{intro.tex}

\section{Background}
\label{sec:background}
\input{background.tex}

\section{Model}
\label{sec:model}
\input{model.tex}

\section{Bound}
\label{sec:bound}
\input{bound.tex}

\section{Checking Admissibility}
\label{sec:checking-admissibility}
\input{admissibility.tex}

\section{Exact Computation}
\label{sec:exact-computation}
\input{computation.tex}

\section{Conclusion}
\label{sec:conclusion}
\input{conclusion.tex}

\bibliographystyle{IEEEtran}
\bibliography{bibliography}


%
\newpage

\appendices






\end{document}

%% file: packages.tex
\ifCLASSINFOpdf
   \usepackage[pdftex]{graphicx}
\else
\fi
\usepackage{amsmath}
\usepackage{amssymb}
\usepackage{qtree}
\usepackage[usenames,dvipsnames]{color}

\newtheorem{definition}{Definition}
\newtheorem{notation}{Notation}
\newtheorem{assumption}{Assumption}
\newtheorem{remark}{Remark}
\newtheorem{example}{Example}

\usepackage{algorithmicx}
\usepackage[noend]{algpseudocode}
\usepackage{url}

\usepackage{epigraph}

%% file: intro.tex
\epigraph{When you have eliminated the impossible, whatever remains, however improbable, must be the truth}{A.C. Doyle, 1890}

Priority Inheritance \cite{PIP} is a widely used protocol for real-time applications involving shared resources with a huge practical and theoretical impact. Its adoption is pervasive in the control and automation industry and in all other domains that rely on real-time systems \cite{Silberschatz}.

The purpose of priority inheritance is to prevent unbounded priority inversion. With respect to other, more efficient protocols proposed in the last years to address the same problem, priority inheritance has a great advantage in its transparency, in the sense that its implementation does not require any information on the tasks involved. It offers, however, a significant drawback, in that there are no known exact methods for computing the blocking time, and the only known method for bounding the blocking time is of exponential complexity \cite{Buttazzo}. 

Blocking time is an essential element in feasibility analysis, which is one key theoretical and practical aspect of real-time systems. While blocking time computation can be done exactly, efficiently and straightforwardly under many other resource access protocols \cite{Buttazzo}, under priority inheritance even \emph{bounding} the blocking time is nontrivial, because there are many possible causes of blocking, and jobs can be blocked multiple times, a phenomenon called chained blocking. The problem becomes particularly intricate when jobs are allowed to hold multiple resources at a time.

In this article \emph{we propose a polynomial method for bounding the blocking time, and an exact, optimally efficient method for blocking time computation under priority inheritance that applies without restrictions on the number of resources each job can hold}. 

We draw from results in operations research and artificial intelligence. 
In particular, we show how the bounding problem can be mapped onto an assignment problem, which is a well-studied problem in operations research. Then we define blocking time computation as a search problem in the space of possible assignments of resources, where the objective is to find the path that induces the worst-case scenario associated with the maximum blocking time. Search can also be seen as a process aimed to eliminate impossible resource assignments, corresponding to inadmissible paths. To that end, we provide a full characterization of the conditions that must be met in order for a resource assignment to be admissible. Moreover, we show that the polynomial bound can be used as an admissible heuristics in the search process. As a consequence, the search method we propose is both \emph{exact} and \emph{maximally efficient}, in the sense that it does not explore branches unnecessarily.

%% file: background.tex
We build on work by Sha, Rajkumar and Lehoczky \cite{PIP}, who proposed and studied two \textit{priority inheritance protocols}: the ``basic'' Priority Inheritance Protocol (PIP), and the Priority Ceiling Protocol (PCP) as a solution to unbounded priority inversion \cite{Buttazzo}. Blocking time is an essential element in feasibility analysis under resource constraints. While PCP's blocking time is perfectly understood, and its computation straightforward, with PIP instead literature only provides upper bounds \cite{Buttazzo}. One such upper bound was proposed by Rajkumar \cite{Rajkumar}. However, using an upper bound for feasibility analysis may be unnecessarily conservative and result in failure to identify perfectly feasible applications with an arbitrarily small processor utilization.

\begin{figure}[ht]
\centering
\includegraphics[width=.7\columnwidth]{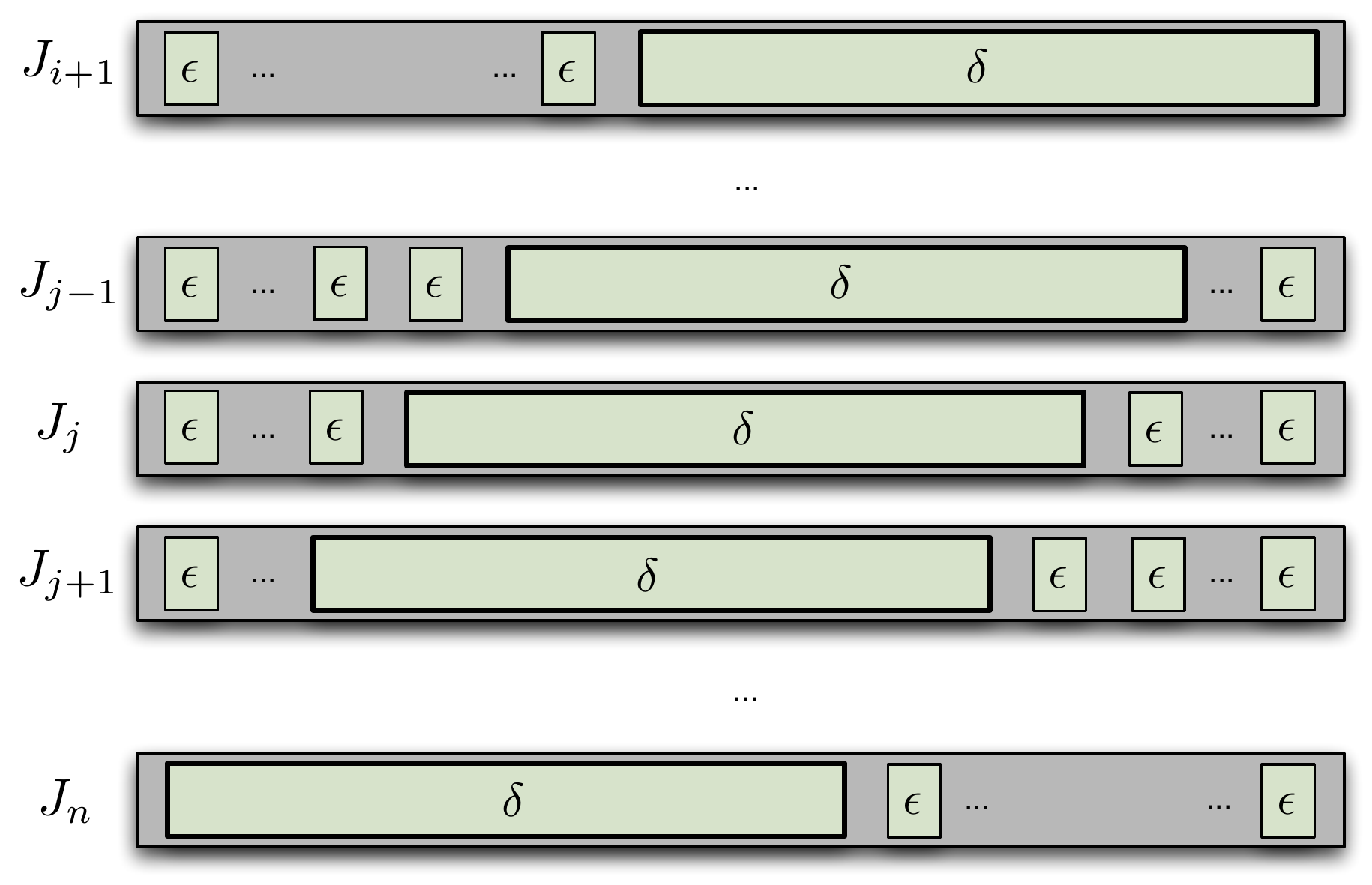}
\caption{Example of resource accesses leading to an overly conservative bound for the PIP blocking time.}
\label{fig:example-loose-bound}
\end{figure}

The following example introduces an application where the upper bound results in an overly conservative blocking time estimation.

\begin{example}
Consider a job $J_i$ with priority $P_i$, which uses $n-i$ resources $\{ R_1, R_2, \dots, R_{n-i} \}$, and a set $\Gamma^i$ of $n-i$ jobs, $\{ J_{i+1}, J_{i+2}, \dots, J_n \}$, with priority $P_{i+1}, P_{i+2}, \dots, P_{n}$, which also use the same resources.
Let the resource associated with a critical section $z_{j,p}$ be $R_p$ (all jobs access resources in the same order). Finally, let the duration of each critical section be: \begin{itemize} \item $\delta$ for $z_{j,n-j+1}$, for all $j$ (i.e., all the sections in the antidiagonal), and \item an arbitrarily small $\epsilon$ in all other cases \end{itemize} as illustrated in Figure~\ref{fig:example-loose-bound}. With this set up, the upper bound obtained by applying Rajkumar's method on $J_i$'s blocking time $B_i$ would be $(n-i)\delta$. However, because of the reasons we will discuss in the next sections, the exact $B_i$ is only $(n-i)\epsilon+\delta$, if $(n-i)$ is odd, or an even smaller $(n-i-1)\epsilon+\delta$, if $(n-i)$ is even. Since $\epsilon$ can be arbitrarily small, the exact value is $n-i$ times smaller than the estimated bound, with self-evident implications on feasibility analysis.
\label{ex:example-loose-bound}
\end{example}

This apparent shortcoming of current feasibility analysis methods and the pervasive use of PIP motivates us to devise an exact procedure for blocking time computation under PIP. In order to do that, we start by introducing   notation, definitions, as well as scheduling model used in literature \cite{PIP,Buttazzo}. For ease of reference we summarize the notation in Table~\ref{tab:notation}.

A \textit{job} is a sequence of instructions that will continuously use the processor until its completion if it is executing alone on the processor. That is, we assume that jobs do not suspend themselves, say for I/O operations.

A \textit{periodic task} is a sequence of jobs of the same type occurring at regular intervals. $J_i$ denotes a job, i.e., an instance of a task $\tau_i$. Each task is assigned a \textit{fixed priority}, and every job of the same task is initially assigned that task's priority. $P_i$ denotes $J_i$'s priority.  We assume that jobs $J_1, J_2, \dots, J_n$ are listed in descending order of priority with $J_1$ having the highest priority, $P_1$. 

If several jobs are eligible to run, the highest-priority job will be run. Jobs with the same priority are executed in FCFS discipline. When a job $J$ is forced to wait for the execution of lower-priority jobs, $J$ is said to be \textit{blocked}.

A binary semaphore guarding a shared resource is denoted by $S$, usually with a subscript, and it provides the $wait$ and $signal$ indivisible operations. 
The $p$-th critical section in $J_j$ is denoted by $z_{j,p}$ and corresponds to the code segment of  $J_j$ between the $p$-th $wait$ operation and its corresponding $signal$ operation. The semaphore that is locked and released by $z_{j,p}$ is denoted by $S_{j,p}$. The resource guarded by $S_{j,p}$ is denoted by $R_{j,p}$.
The \textit{duration} of a critical section $z_{j,p}$, denoted $d_{j,p}$, is the time to execute $z_{j,p}$ when $J_j$ executes on the processor alone.
A job $J_i$ is said to be  blocked by the critical section $z_{j,p}$ of job $J_j$ if $i<j$ and $J_i$ has to wait for $J_j$ to exit $z_{j,p}$ in order to continue execution.
The sequence of all critical sections of a job $J_j$ is denoted by $\beta_{j} = \langle \dots, z_{j,p}, \dots \rangle$. 

\begin{table}[!bt]
\begin{center}
\begin{tabular}{|c|p{.8\columnwidth}|}
    \hline
    Symbol & Meaning \\
    \hline
    $\tau_j$   			& $j$-th \emph{periodic task} \\
    $P_j$  		& the \emph{priority} associated with $\tau_j$ 	\\
    $J_j$   			& $j$-th \emph{job}: an instance of $J_j$ \\
   $\Gamma$ 		& a set of jobs (\emph{application}) \\
   $\Gamma^i$ 		& the set of all jobs in $\Gamma$ that can block $J_i$ \\
$\Gamma_N^i$ & the set of jobs that can block $J_i$ when jobs can hold more than one resource at a time \\
    $z_{j,p}$  		& $p$-th \emph{critical section} of $J_j$, corresponding to the code segment of $J_j$ between the $p$-th wait operation and its corresponding signal operation	\\
    $z_{j,p} \subset z_{j,s}$   			& $z_{j,p}$ is \emph{entirely contained} in $z_{j,s}$  \\
    $\beta_j$ 			& the sequence of all critical sections of a job $J_j$: $\langle \dots, z_{j,p}, \dots \rangle$ 	\\
$\beta_{j}(\mathcal{\hat{R}})$ & \emph{maximal sequence} with respect to $\mathcal{\hat{R}}$: $\langle z_{j,p} \in \beta_{j}| z_{j,p} \textrm{ is maximal with respect to } \mathcal{\hat{R}}\rangle$ \\
$\mathcal{Z}$ & a chain of critical sections, or \emph{z-chain}: $\langle \dots, z_{j,k}, \dots \rangle$ \\
    $d_{j,p}$  		& $z_{j,p}$'s \emph{duration}	\\
	$d(\mathcal{Z})$ & $\mathcal{Z}$'s duration \\
    $S_{j,p}$			& the \emph{semaphore} associated with $z_{j,p}$ \\
    $R_{j,p}$			& the \emph{resource} guarded by $S_{j,p}$ \\
   $\mathcal{R}^i$ 		& the set of all resources whose semaphores can block $J_i$ when each job can hold at most one resource at a time \\
   $\mathcal{R}$	& the set of all resources used by jobs in $\Gamma$ \\
$\mathcal{R}_N^i$ & the set of all resources whose semaphores can block $J_i$ when jobs can hold more than one resource at a time \\
$in(J_i, z_{j,p},\mathcal{\hat{R}})$ & set induced by ($J_i, z_{j,p}$) from $\mathcal{\hat{R}}$ \\
$in(J_i, \mathcal{Z})$ & set induced by ($J_i,\mathcal{Z}$) \\
    \hline  \end{tabular}
\end{center}
\caption{Notation}
\label{tab:notation}
\end{table}

As in \cite{PIP}, we use a simplified scheduling model, as defined by the following assumptions.

\begin{assumption} All the tasks are periodic. 
\end{assumption}
\begin{assumption} Access to shared resources is regulated by the basic priority inheritance protocol defined in \cite{PIP}. In particular, when a job $J_i$ blocks one or more higher-priority jobs, it temporarily assumes the highest priority of the blocked jobs.\footnote{For a formal definition of the protocol see \cite{PIP}.}
\end{assumption}
\begin{assumption} Each job in a periodic task has deterministic execution times for both its critical and noncritical sections and it does not synchronize with external events, i.e., a job will execute to its completion when it is the only job in the system.
\end{assumption}

The last assumption implies that the sequence of operations on semaphores by each individual job is known, and that the worst-case execution time of each critical section is also known.\footnote{Blocking time analysis typically considers only the \emph{longest} critical sections \cite{PIP}. However, an \emph{exact} computation of the worst-case blocking time under PIP requires more information.} In particular, we will describe each job by the sequence and length of its critical sections. 

Current work on blocking time analysis under PIP typically assumes that a job can hold only a resource at a time. We instead accept that jobs can hold multiple shared resources at the same time. However, following a well-established convention \cite{Buttazzo}, we assume proper nesting of critical sections.
We shall write $z_{j,p} \subset z_{j,s}$, or equivalently $z_{j,s} \supset z_{j,p}$, if a critical section $z_{j,p}$ is \emph{entirely contained} in $z_{j,s}$ \cite{PIP}. 

\begin{assumption}
We assume that critical sections are \emph{properly nested}. That is, given any pair of critical sections $z_{j,s}$ and $z_{j,p}$, if $s<p$, then either  $z_{j,p} \subset z_{j,s}$, or $z_{j,s} \cap z_{j,p} = \emptyset$. 
Moreover, we assume that a semaphore may be locked at most once in a single nested critical section, so $z_{j,s} \supset z_{j,p} \Rightarrow R_{j,s} \neq R_{j,p}$ \cite{PIP}.
\end{assumption}

Finally, we assume that resources are properly released.

\begin{assumption}
Each job releases before terminating any resource it holds.
\end{assumption}

When convenient, we will use square brackets to denote critical sections, indicating in the brackets the name of the associated resources and the duration of the section.

\begin{example}
The following notation:
\begin{description}
\item[$J_1$] $[R_2: 3 ~[ R_1: 1]]$
\item[$J_2$] $[R_1: 3] ~ [ R_1: 4]$
\end{description}
describes a set of two jobs: $J_1$ with two critical sections, $z_{1,1}$ and $z_{1,2}$, and $J_2$ with two critical sections, $z_{2,1}$ and $z_{2,2}$. The duration of $z_{1,1}$ is $d_{1,1}=3$, and the resource associated with $z_{1,1}$ is $R_{1,1}=R_2$, guarded by semaphore $S_2$. $z_{1,2}$ is entirely contained in $z_{1,1}$, whereas $z_{2,2}$ follows $z_{2,1}$.
\label{ex:square-brackets}
\end{example}

We will call an ordered sequence of critical sections a \textbf{z-chain}, denoted as $\mathcal{Z} = \langle \dots, z_{j,p}, \dots \rangle$. The \emph{duration} of a z-chain, denoted $d(\mathcal{Z})$, is the sum of the durations of its elements:  $$d(\mathcal{Z}) = \sum_{z_{j,p}\in\mathcal{Z}} d_{j,p}$$

%% file: model.tex
In this section we will identify and define all the elements that are necessary for an analysis of the blocking time computation under PIP.

Consider an application $\Gamma = \{J_{1}, \dots,  J_i, \dots, J_n\}$ and a set of resources  $\mathcal{R} = \{R_1, R_2, \dots, R_m\}$, each guarded by a distinct binary semaphore.

It is a known fact that, if each job can hold at most one resource at a time, $\mathcal{R}^i$ includes all and only the resources used \textit{both} by jobs with priority lower than $P_i$ and by jobs with priority higher than or equal to $P_i$ \cite{PIP}. We will use $\mathcal{R}^i$ to denote the set of resources whose semaphores \textbf{can cause blocking} to $J_i$ if each job can hold at most one resource a time: $$\mathcal{R}^i = \{ R \in \mathcal{R} | \exists z_{j,p} \in \beta_j, z_{k,q} \in \beta_k,  k \leq i, i < j | R_{k,q}=R_{j,p}=R \}.$$

Accordingly, we will use $\Gamma^i$ to denote the set of all jobs that \textbf{can block} $J_i$, if each job can hold at most one resource at a time. In particular, $\Gamma^i$ includes all and only the jobs with priority lower than $P_i$ that use resources belonging to $\mathcal{R}^i$ \cite{PIP}: $$\Gamma^i = \{ J_j \in \Gamma | j>i, \exists z_{j,p} \in \beta_j | R_{j,p} \in \mathcal{R}^i \}.$$

The fact that critical sections can be nested, properly or otherwise, introduces the threat of \textbf{deadlock}.\footnote{Deadlock is not an issue when all sections are disjoint, because a deadlock requires the occurrence of the \emph{hold-and-wait} condition \cite{Silberschatz}, which cannot occur if all sections are disjoint.} Clearly, deadlocks must be prevented in real-time applications.
A common way to do so is by preventing a necessary condition for deadlock, known as \textit{circular wait}, in particular by imposing a strict order on resource acquisitions. Checking that a given application respects such a strict order is trivial.\footnote{One possibility is to map resources onto vertices of a directed graph, and the ``entirely contained'' relation onto edges between vertices. Then one can use a linear-time method such as Tarjan's strongly connected components algorithm \cite{Tarjan} to verify that the graph has no strongly connected subgraphs with more than one vertex, i.e., the graph is a directed acyclic graph. If the graph is acyclic, deadlock cannot occur. 
} 
We will thus assume that deadlock is prevented by some external means, and in particular that semaphores are accessed in an order consistent with a predefined acyclical order \cite{PIP}:
\begin{assumption}
We assume that the $\subset$ relation defined over nested critical sections induces a partial order over resources. 
\end{assumption}

Nesting also introduces a new phenomenon, called \textbf{transitive priority inheritance}~\cite{Buttazzo}. In particular, if a job $J_i$ is blocked by a job $J_j$, and $J_j$ is blocked by a third job $J_k$, then $J_k$ inherits $J_i$'s priority via $J_j$.\footnote{It is well-known that transitive priority inheritance  is only possible in the presence of nested sections.}

An effect of transitive priority inheritance is the extension of the set of resources that can cause blocking to $J_i$. 
In the absence of nested sections, when each job can hold at most one resource at a time, a resource can block $J_i$ only if its ceiling is at least $P_i$, and it is used by a job with a priority lower than $P_i$. This no longer holds. In the presence of nested sections, because of transitive inheritance, a job can inherit a priority higher than that of the job it's blocking. Therefore, a resource can cause blocking to $J_i$ even if its ceiling is \textit{lower} than $P_i$, but higher than or equal to the priority of the jobs \textit{that can inherit} a priority greater than or equal to $P_i$. The set of jobs that can block $J_i$ is thus, in general, a superset of $\Gamma^i$.

\begin{figure*}[t]
\centering
\includegraphics[width=\columnwidth]{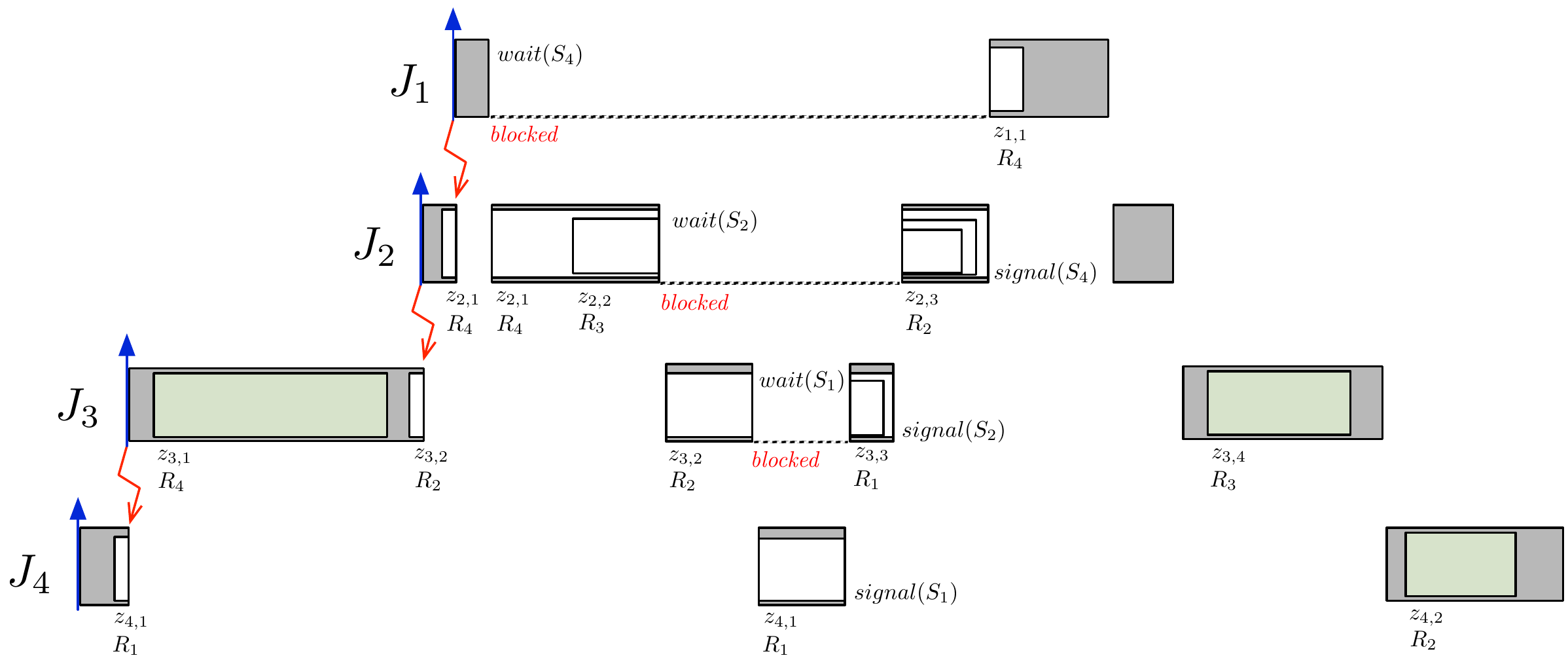}
\caption{A possible scheduling of the application described in Example~\ref{ex:nested-sections}.}
\label{fig:example-nested-sections}
\end{figure*}

\begin{example}
Let us consider $\Gamma = \{ J_1, J_2, J_3, J_4 \}$. Let jobs in $\Gamma$ access a set of shared resources $\mathcal{R} = \{ R_1, R_2, R_3, R_4 \}$, in the following way:
\begin{description}
\item[$J_1$] $[R_4:1]$
\item[$J_2$] $[R_4:6 ~[R_3: 4 ~[R_2:2]]]$
\item[$J_3$] $[R_4:10] ~[R_2:3 ~[R_1:1]] ~[R_3:5]$
\item[$J_4$] $[R_1:2] ~[R_2:4]$
\end{description}
These jobs define the following sequences of critical sections: $\beta_1 = \langle z_{1,1} \rangle$,  $\beta_2 = \langle z_{2,1}, z_{2,2}, z_{2,3} \rangle$,  $\beta_3 = \langle z_{3,1}, z_{3,2}, z_{3,3}, z_{3,4} \rangle$, and $\beta_4 = \langle z_{4,1}, z_{4,2} \rangle$. We observe that $z_{2,3} \subset z_{2,2}$, $z_{2,2} \subset z_{2,1}$, and $z_{3,3} \subset z_{3,2}$, which together with the fact that $R_{2,3}=R_2$, $R_{2,2}=R_3$, $R_{2,1}=R_4$, $R_{3,3}=R_1$, and $R_{3,2}=R_2$, induces a resource ordering $R_1 < R_2 < R_3 < R_4$, thus $\Gamma$ is deadlock-free. 

We have $\mathcal{R}^1=\{R_4\}$ and  $\Gamma^1 = \{ J_2, J_3 \}$, so if the critical sections were all disjoint, $J_4$ could not possibly cause blocking to $J_1$, and we would have $B_1=d_{3,1}=10$. 

However, let us consider the sequence of events illustrated in Figure~\ref{fig:example-nested-sections}, where $J_3$ is released as soon as $J_4$ acquires $S_1$ and enters $z_{4,1}$, $J_2$ is released as soon as $J_3$ acquires $S_2$ and enters $z_{3,2}$, and finally $J_1$ is released as soon as $J_2$ acquires $S_4$ and executes $z_{2,1}$. In that case, as soon as $J_1$ attempts to acquire $S_4$ (the semaphore guarding $z_{1,1}$ as well as $z_{2,1}$), $J_1$ will be blocked for the duration of the whole z-chain $\mathcal{Z} = \langle z_{4,1}, z_{3,2}, z_{2,1} \rangle$, that is, for 11 units of time. Interestingly, $\mathcal{Z}$ involves sections that are not directly associated with   $\mathcal{R}^1$ and $\Gamma^1$: 
$J_4$ (not in $\Gamma^1$) has a section that belongs to $\mathcal{Z}$, $z_{4,1}$, which uses $R_{4,1}=R_1$, also not in $\mathcal{R}^1$; however, $R_1$ contributes to blocking because $R_1=R_{3,3}$ and $z_{3,3} \subset z_{3,2}$, and in turn $R_{3,2}=R_2=R_{2,3}$ and $z_{2,3} \subset z_{2,1}$, with, finally, $R_{2,1}=R_4\in\mathcal{R}^1$.
In the end, the set of resources that cause blocking to $J_1$ in this example is
$\{R_1, R_2, R_4\} \supset \mathcal{R}^1$, and the set of jobs that block $J_1$ is $\{ J_2, J_3, J_ 4 \} \supset \Gamma^1$.
\label{ex:nested-sections}
\end{example}

The example above motivates the introduction of the set
$\mathcal{R}_N^i \supseteq \mathcal{R}^i$, which includes all and only the resources whose semaphores can cause blocking to $J_i$ when nested sections are allowed. Accordingly, $\Gamma_N^i \supseteq \Gamma^i$ denotes the set of jobs that can block $J_i$ when nested sections are allowed.

In particular, $\mathcal{R}_N^i$  includes all and only the resources used both by jobs with priority lower than $P_i$, and by jobs that have \textit{or can inherit} a priority equal to or greater than $P_i$ (due to transitive priority inheritance).  In order to characterize $\mathcal{R}_N^i$ and $\Gamma_N^i$ we need to delve a bit deeper into such a phenomenon.

Transitive priority inheritance requires three distinct jobs, $J_i$, $J_j$, and $J_k$. If these are the only jobs, then in order for $J_k$ to inherit $P_i$ through $J_j$, the following conditions must hold: (1) $J_j$ defines two critical sections, $z_{j,p}$ and $z_{j,q}$, such that $z_{j,p} \supset z_{j,q}$, (2) $R_{j,p}$ is shared with $J_i$ and (3) $R_{j,q}$ is shared with $J_k$. 

More in general, we can say that a job $J_{k}$ can cause blocking to $J_{i}$ either because, independently of nested sections, $J_{k} \in \Gamma^i$, or because the following conditions hold: (1) a \emph{third} job $J_j$, \emph{with priority lower than} $P_i$, defines two critical sections, $z_{j,p}$ and $z_{j,q}$, such that $z_{j,p} \supset z_{j,q}$, (2) the resource associated with the \emph{outer} section, $R_{j,p}$, is \emph{a resource that can cause blocking to} $J_i$, and (3) $J_k$ defines a critical section that uses $R_{j,q}$. Under such conditions, $R_{j,q}$  can cause blocking to $J_i$. Notice that the blocking in question does not depend on $J_j$ and $J_k$'s relative priority, as long as $J_k$'s priority $P_k$ is lower than $P_i$, and $J_j$ is other than $J_k$.
We then obtain the following characterization: 
$$\mathcal{R}_N^i = \mathcal{R}^i \cup \{  R_{j,q} | \exists z_{j,q} \subset z_{j,p}, R_{j,p} \in \mathcal{R}_N^i, \exists z_{k,r} | R_{k,r}=R_{j,q}, k>i, k\neq j \}.$$

Accordingly, $\Gamma_N^i$ includes all and only the jobs with priority lower than $P_i$, that use resources belonging to $\mathcal{R}_N^i$:
$$\Gamma_N^i = \{ J_j \in \Gamma | j>i, \exists z_{j,k}, R_{j,k} \in \mathcal{R}_N^i \}.$$

\begin{example}
[continued from~\ref{ex:nested-sections}] We have $\mathcal{R}_N^1 =\{ R_4 \} \cup \{ R_3, R_2, R_1 \}$, $\Gamma_N^1 = \{ J_2, J_3, J_4 \}$, $\mathcal{R}_N^2 =\{ R_2, R_3, R_4 \} \cup \{R_1\}$, $\Gamma_N^2 = \{ J_3, J_4\}$, $\mathcal{R}_N^3 = \mathcal{R}^3 =\{ R_1, R_2 \}$, and $\Gamma_N^3 = \{  J_4\}$.
\label{ex:nested-sections-extended}
\end{example}

$\mathcal{R}_N^i$ defines the resources  that \emph{in principle} could block $J_i$. However, blocking depends on the schedule, and not all schedules are possible. To illustrate, consider the following example.

\begin{example}
[continued from~\ref{ex:nested-sections-extended}] From previous analysis we know that $\mathcal{Z} = \langle z_{4,1}, z_{3,2}, z_{2,1} \rangle$ corresponds to a possible schedule (illustrated in Figure~\ref{fig:example-nested-sections}), yielding an overall blocking time for $J_1$ of 11 time units. $\mathcal{Z}$ corresponds to the following allocation of resources in $\mathcal{R}_N^1$ to jobs in $\Gamma_N^1$: $R_1/J_4$, $R_2/J_3$, and $R_4/J_2$.
Let us now consider a different z-chain $\mathcal{Z}'$,  also involving three different resources/jobs in $\mathcal{R}_N^i$/$\Gamma_N^i$: $\mathcal{Z}' = \langle z_{4,2}, z_{3,4}, z_{2,1} \rangle$, yielding a total duration $d(\mathcal{Z}')=4+5+6=15$. $\mathcal{Z}'$ corresponds to the following allocation of resources in $\mathcal{R}_N^1$ to jobs in $\Gamma_N^1$: $R_2/J_4$, $R_3/J_3$, and $R_4/J_2$. The jobs and resources are the same as before, but unlike $\mathcal{Z}$, $\mathcal{Z}'$ describes an impossible schedule. Indeed, $J_3$ may not obtain access to $z_{3,4}$ while $J_4$ holds $R_2$, because in order to reach $z_{3,4}$, $J_3$ should cross $z_{3,2}$, meaning acquiring (and then releasing) $R_2$. 

Moreover, if we consider other possible allocations that could cause blocking to $J_1$, we notice that each allocation where $J_3$ holds $R_4$ would inhibit any possible contribution of $R_1$, $R_2$, and $R_3$ towards blocking $J_1$. 
As a matter of fact, $R_2$ and $R_3$ belong to $\mathcal{R}_N^1$ only by virtue of $J_2$ potentially holding $R_4$, and $R_1$ belongs to $\mathcal{R}_N^1$ only by virtue of $J_3$ potentially holding $R_2$ even as $J_2$ holds $R_4$. 

As a result, the only possible allocation where all the resources in $\mathcal{R}_N^1$ play a role towards $B_1$ is that corresponding to $\mathcal{Z}$ in Example~\ref{ex:nested-sections}. Another possible allocation of resources yielding the same duration would be $R_3/J_3$, $R_4/J_2$, and in that case $J_4$ may not hold any resource. Other possible allocations result in shorter z-chains, therefore the duration of the longest z-chain for this application, corresponding to a possible schedule, is $B_1=11$ units.
\label{ex:nested-sections-z-chains}
\end{example}

In general, whether a resource may or may not belong to a z-chain corresponding to an admissible schedule depends on the other resources in the same z-chain. We shall thus introduce the notion of a \emph{induced} resource set.
This will enable us define an iterative characterization of $\mathcal{R}_N^i$ equivalent to the recursive one given earlier. The idea is to obtain $\mathcal{R}_N^i$ by initially computing $\mathcal{R}^i$ and then iteratively applying the definition of induced set until a fix point is reached. But before we go there, we need to introduce the notion of maximality with respect to a set of resource.

\begin{definition}[Maximal section]
Given a set $\mathcal{\hat{R}}$ of resources, a section $z_{j,p}$ is \textbf{maximal with respect to} $\mathcal{\hat{R}}$ if and only if $R_{j,p} \in \mathcal{\hat{R}}$ and $\nexists z_{j,s} \supset z_{j,p} | R_{j,s} \in \mathcal{\hat{R}}$.
\label{def:maximal-section}
\end{definition}

\begin{definition}[Maximal sequence]
Given a set $\mathcal{\hat{R}}$ of resources and a sequence $\beta_{j}$, the corresponding \textbf{maximal sequence with respect to} $\mathcal{\hat{R}}$, denoted  $\beta_{j}(\mathcal{\hat{R}})$, is the sequence of  sections in $\beta_{j}$ that are maximal with respect to $\mathcal{\hat{R}}$: $\beta_{j}(\mathcal{\hat{R}}) = \langle z_{j,p} \in \beta_{j}| z_{j,p} \textit{ is maximal with respect to } \mathcal{\hat{R}}\rangle$. 
\label{def:maximal-sequence}
\end{definition}

\begin{definition}[Induced set]
Let $\mathcal{\hat{R}}$ be a set of resources, $J_i$ a job, and $z_{j,p}$ a \emph{maximal} section with respect to $\mathcal{\hat{R}}$, for some $j>i$. 
The \textbf{set induced by} ($J_i, z_{j,p}$) \textbf{from} $\mathcal{\hat{R}}$, denoted $in(J_i, z_{j,p},\mathcal{\hat{R}})$, is 
 the set  of resources $R_{j,q}$ that 
 (1) are associated with a critical section $z_{j,q}$ \emph{contained in} $z_{j,p}$, 
 (2) \emph{do not belong} to $\mathcal{\hat{R}}$, and
 (3) are associated with a critical section belonging to a job \emph{other than} $J_j$ and with a priority \emph{lower than} $P_i$:
 $$in(J_i, z_{j,p},\mathcal{\hat{R}}) = \{ R_{j,q}  | z_{j,q} \subset z_{j,p}, R_{j,q} \in \mathcal{R} \setminus \mathcal{\hat{R}}, \exists z_{k,r}| R_{k,r}=R_{j,q}, k>i, k\neq j \}.$$
\label{def:induced-sets}
\end{definition}

\begin{example}
[continued from~\ref{ex:nested-sections-z-chains}] Consider $J_1$, $\mathcal{R}^1$, and $z_{2,1}$, which is maximal with respect to $\mathcal{R}^1$. We have $in(J_1, z_{2,1},\{R_4\}) = \{ R_2, R_3 \}$. Indeed, if $J_2$ enters $z_{2,1}$ while $R_2$ or $R_3$ are held by other jobs, $J_2$ will not be able to complete its execution of $z_{2,1}$ and thus release $R_4=R_{2,1}$ until it can get hold of $R_2$ and $R_3$ as well.
\label{ex:nested-sections-induced-set}
\end{example}

Induced sets can be used to compute $\mathcal{R}_N^i$. The straightforward way to do that is to initially set $\mathcal{R}_N^i=\mathcal{R}^i$ and then apply the induction operator until a fix point is reached.
Such a method, encoded by function \textsc{Relevant-Resources} in Figure~\ref{algo:RNi}, will necessarily reach a fix point, because  $\mathcal{R}_N^i$ is a monotonically growing set of resources, and resources are finite. Moreover, its complexity is bound by the number of resources outside of $\mathcal{R}^i$ times the number of critical sections in jobs with a priority lower than $P_i$.

\begin{figure}
\begin{algorithmic} [1]
\Function{Relevant-Resources}{$\Gamma,i$} 
\State $\mathcal{R}_N^i \gets \mathcal{R}^i$
\While{$\mathcal{R}_N^i \subset \mathcal{R}$ \textbf{and} $\exists$ \emph{maximal} $z_{j,k}$ \emph{wrt} $\mathcal{R}_N^i ~|~ j>i$ \textbf{and} $in(J_i, z_{j,k},\mathcal{R}_N^i) \neq \emptyset$}
\State $\mathcal{R}_N^i \gets \mathcal{R}_N^i \cup in(J_i, z_{j,k},\mathcal{R}_N^i)$
\EndWhile
\State \Return $\mathcal{R}_N^i$
\EndFunction
\end{algorithmic}
\caption{Iterative computation of $\mathcal{R}_N^i $}
\label{algo:RNi}
\end{figure}

\begin{example}
[continued from~\ref{ex:nested-sections-induced-set}]  
${\mathcal{R}_N^1}^{(0)}=\mathcal{R}^1=\{R_4\} \subset \mathcal{R}$. 
$in(J_1,z_{2,1},{\mathcal{R}_N^1}^{(0)})=\{R_2, R_3\}$.
${\mathcal{R}_N^1}^{(1)} = {\mathcal{R}_N^1}^{(0)} \cup in(J_1,z_{2,1},{\mathcal{R}_N^1}^{(0)}) = \{R_2, R_3, R_4\} \subset \mathcal{R}$.
Maximal sections of $J_2$, $J_3$, and $J_4$ with respect to ${\mathcal{R}_N^1}^{(1)}$: $z_{2,1}$, $z_{3,1}$, $z_{3,2}$, $z_{3,4}$, and $z_{4,2}$.
$in(J_1,z_{2,1},{\mathcal{R}_N^1}^{(1)})=\emptyset$.
$in(J_1,z_{3,1},{\mathcal{R}_N^1}^{(1)})=\emptyset$.
$in(J_1,z_{3,2},{\mathcal{R}_N^1}^{(1)})=\{R_1\}$.
${\mathcal{R}_N^1}^{(2)} = {\mathcal{R}_N^1}^{(1)} \cup in(J_1,z_{3,2},{\mathcal{R}_N^1}^{(1)}) = \mathcal{R}$ (fix point).
\label{ex:induced-set-fixpoint}
\end{example}

Definition~\ref{def:induced-sets} applies a single section, but we can  extend it to z-chains.

\begin{definition}
Let $\mathcal{Z}$ be a z-chain of sections that can cause blocking to $J_i$. 
The set induced by ($J_i, \mathcal{Z}$), denoted $in(J_i, \mathcal{Z})$, is defined as 
$\mathcal{R}^i \cup \bigcup_{z \in \mathcal{Z}} in(J_i,z,\mathcal{R}^i)$.
\end{definition}

We are now ready to characterize all the possible cases of blocking using the notion of \textbf{admissibility} and its necessary condition, \textbf{induction compatibility}.  Intuitively, a z-chain  $\mathcal{Z}$ is \emph{induction compatible} if each resource associated to sections in  $\mathcal{Z}$ contributes to blocking $J_i$, given the other elements in  $\mathcal{Z}$, whereas it is \emph{admissible} if it is induction compatible and corresponds to a possible schedule. In that case, $\mathcal{Z}$ describes a possible sequence of job activations leading to a situation where at a given time each relevant job executes inside its corresponding section in $\mathcal{Z}$, whereby the total blocking $J_i$ is subject to is $B_i=d(\mathcal{Z})$. If, otherwise, $\mathcal{Z}$ is inadmissible, $\mathcal{Z}$ cannot cause a blocking $B_i=d(\mathcal{Z})$, because it is impossible to schedule jobs so as to have at any given time all relevant job executing inside their corresponding section in $\mathcal{Z}$.

\begin{definition}[Induction compatibility]
Consider a job $J_i$ and a z-chain $\mathcal{Z}$ of sections belonging to all-different tasks and associated with all-different resources. Then a section $z_{j,p} \in \mathcal{Z}$ is \emph{induction compatible} if either $R_{j,p} \in \mathcal{R}^i$ or $\exists  z_{k,q}, z_{k,r} \in \beta_k$ for some $k \neq j$ such that $z_{k,q} \in \mathcal{Z}, z_{k,q} \supset z_{k,r}, R_{k,r}=R_{j,p}$, and $z_{k,q}$ is induction compatible.
\end{definition}

\begin{example}
[continued from~\ref{ex:induced-set-fixpoint}] 
Consider $\mathcal{Z}' = \langle z_{4,2}, z_{3,4}, z_{2,1} \rangle$ from Example~\ref{ex:nested-sections-z-chains}. $z_{2,1}$ is induction compatible because $R_{2,1} \in \mathcal{R}^1$, while $z_{3,4}$ and $z_{4,2}$ are induction compatible because there are two sections contained in $z_{2,1}$ and associated with $R_{3,4}$ and $R_{4,2}$. However, as we know from Example~\ref{ex:nested-sections-z-chains}, $\mathcal{Z}'$ models an impossible  schedule. Consider now $\mathcal{Z}'' = \langle z_{3,2} \rangle$, which represents a perfectly possible job scheduling, where $J_3$ has reached $z_{3,2}$ and is holding $R_1$ and $R_2$. $z_{3,2}$ alone is not induction compatible, because $R_{3,2} = R_2 \not\in \mathcal{R}^1$ and there is no other induction compatible section in $\mathcal{Z}''$ which contains a section associated with $R_2$. Indeed, there is no reason why $\mathcal{Z}''$ should cause any blocking to $J_1$.
\label{ex:induction-compatible}
\end{example}

\begin{figure}[t]
\centering
\includegraphics[width=\columnwidth]{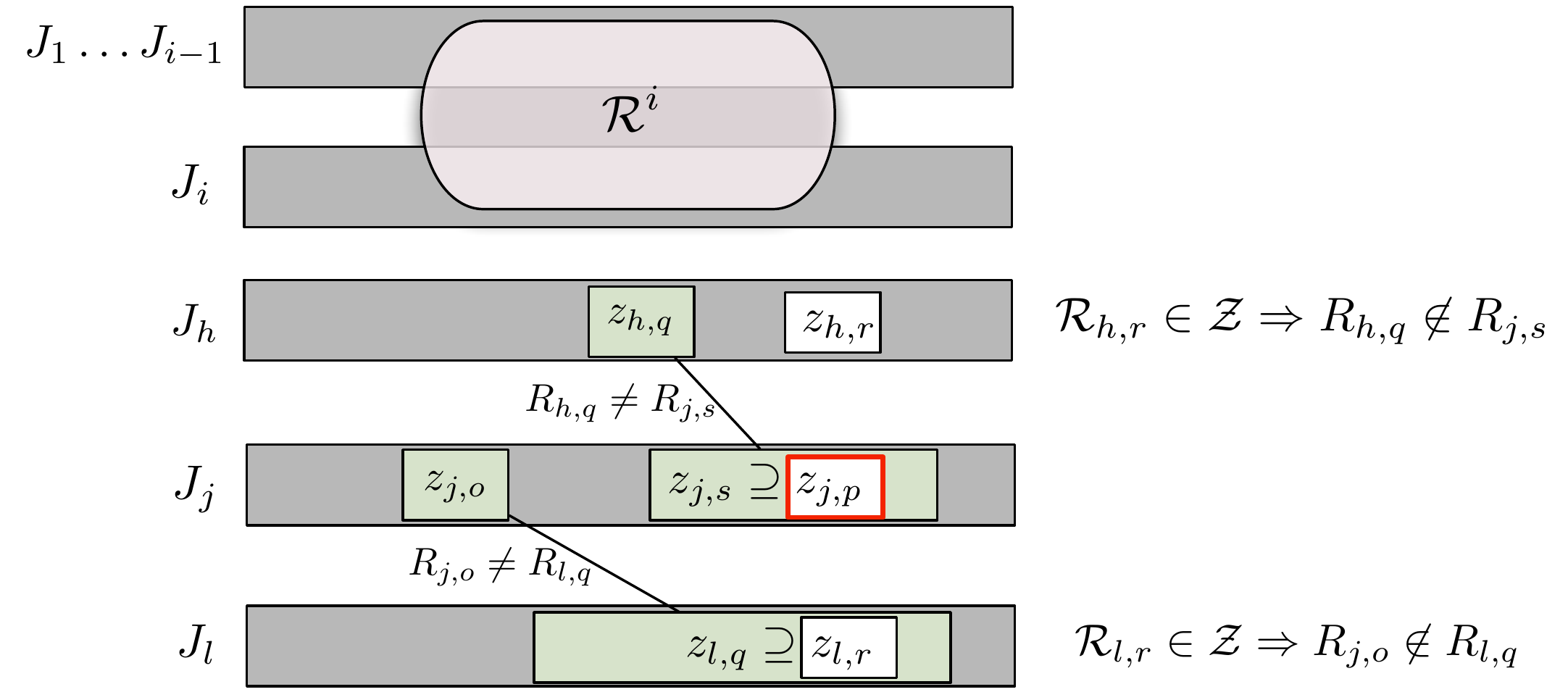}
\caption{Obstructing sections rendering $z_{j,p}$ inadmissible (Definition~\ref{def:admissibility}).}
\label{fig:admissibility}
\end{figure}

Admissibility uses and extends induction compatibility by laying out all the constraints that must be satisfied in order for a z-chain $\mathcal{Z}$ of duration $d$ to cause a blocking $B_i=d$ to a job $J_i$. Admissibility is defined by induction. Figure~\ref{fig:admissibility}  is meant as a reference to clarify the notation used in some constraints (FHO and FLO).

\begin{definition}[Admissibility]
Admissibility  is defined with respect to a job $J_i$ by induction:
\begin{itemize}
\item
The empty chain $\langle \rangle$ is admissible with respect to any job $J_i$.

\item
A non-empty z-chain $\mathcal{Z}'= \mathcal{Z} + \langle z_{j,p} \rangle$ is admissible with respect to a job $J_i$ if and only if $\mathcal{Z}$ is admissible \emph{and} $z_{j,p}$ is an \emph{admissible extension} to $\mathcal{Z}$ with respect to $J_i$. 

\item
A section $z_{j,p}$ is an \emph{admissible extension} to $\mathcal{Z}$ with respect to $J_i$ if an only if it satisfies \emph{all} the following conditions:
\begin{enumerate}

\item[\textbf{NBJ}] (\textit{Novelty of Blocking Job}):
$J_j$ is \textit{a new job}: $\beta_{j} \cap \mathcal{Z} = \emptyset$; 

\item[\textbf{NBR}] (\textit{Novelty of Blocking Resource}): 
$z_{j,p}$ is associated with \textit{a new resource}: $$\nexists z_{k,r} \in \mathcal{Z} | R_{k,r} = R_{j,p};$$

\item[\textbf{LSM}] (\textit{Limited-Scope Maximality}):
$z_{j,p}$ is \textit{maximal} with respect to $in(J_i,\mathcal{Z})$: $$R_{j,p} \in in(J_i,\mathcal{Z}) \textrm{ and } \nexists z_{j,s} \supset z_{j,p} | R_{j,s} \in in(J_i,\mathcal{Z});$$

\item[\textbf{FHO}] (\textit{Freedom from Higher-priority job Obstruction}):
$R_{j,p}$ is not associated with, or contained in a section $z_{j,s}$ associated with, a section $z_{h,q}$ of a \textit{higher} priority job $J_h$ that precedes a section $z_{h,r} \in \mathcal{Z}$: $$\nexists z_{h,r} \in \mathcal{Z}, h<j,  z_{j,s} \in \beta_{j}, z_{j,s} \supseteq z_{j,p},  z_{h,q} \in \beta_{h}, q<r | R_{h,q} = R_{j,s};$$

\item[\textbf{FLO}] (\textit{Freedom from Lower-priority job Obstruction}):
$z_{j,p}$ is not preceded by a section $z_{j,o}$ associated with a resource associated with a section $z_{l,r} \in \mathcal{Z}$, or with a section $z_{l,q}$ containing a section $z_{l,r} \in \mathcal{Z}$, of a \textit{lower} priority job $J_l$: $$\nexists z_{l,r} \in \mathcal{Z}, j<l, z_{l,q} \in \beta_l, z_{l,q} \supseteq z_{l,r}, z_{j,o} \in \beta_j, s<p| R_{j,o} =R_{l,q}.$$

\end{enumerate}
\end{itemize} 
\label{def:admissibility}
\end{definition}

These definitions provide a complete characterization of the conditions for blocking in the absence of nested sections. In particular, \textbf{NBJ} and \textbf{NBR} are known from literature \cite{Buttazzo}: it should be self-evident that in order for $J_i$ to be blocked by two different critical sections of tasks in $\Gamma$, these critical section must refer to different resources and belong to different jobs. 

\textbf{LSM} instead reflects the following observation: if $\mathcal{Z}$ already contains a section associated with a resource $R$, then any other section \emph{contained} in a section associated with $R$ cannot be an admissible extension to $\mathcal{Z}$, since that section could not possibly be reached (the job would be blocked before). Therefore, we are only interested in maximal sections. Moreover, only considering resources belonging to the set induced by $\mathcal{Z}$ ensures induction compatibility. Thus \emph{limited scope} maximality--the scope being limited to $in(J_i,\mathcal{Z})$--rather than just maximality.

Finally, \textbf{FHO} an \textbf{FLO} are reachability conditions. On the one hand, the sections that already belong to $\mathcal{Z}$ should remain reachable, therefore new sections that extend $\mathcal{Z}$ should not obstruct them. On the other hand, these new extensions to $\mathcal{Z}$ must themselves be reachable. Notice that, because jobs can hold multiple resources at the same time, a resource can be either directly associated with a section $z$, or it can be associated with a section that contains $z$, and will thus be allocated to the job that executes $z$. In particular, FHO stipulates that if a section $z_{j,p}$ is added to a chain that contains higher-priority sections, the latter must still be reachable, whereas FLO stipulates that $z_{j,p}$ must itself be reachable in spite of lower-priority sections that may  already be in $\mathcal{Z}$. Reachability is obstructed by sections in the higher-priority job that precede the higher-priority section and are associated with resources that are also associated with the lower-priority section, directly or otherwise.

It is worthwhile noticing that any 1-element z-chain is admissible if and only if its element is maximal with respect to $\mathcal{R}^i$. Any z-chain composed of first-only sections ($\forall z_{k,p} \in \mathcal{Z}, p=1$) satisfies FHO and FLO (as well as, trivially, NBJ), and is therefore admissible if and only if it satisfies NBR and is induction-compatible.

\subsubsection*{The bottom line}

The model we introduce provides a complete characterization of the sequences of critical sections that can block a job. Given such a model, we propose the following methodology for computing the blocking time:

\begin{enumerate}

\item Because nested sections under PIP introduce the risk of deadlock, the  first step is to establish that semaphores are accessed in an order consistent with a predefined acyclical order. If that is not the case, the blocking time is infinity. This can be done in \emph{linear} time.

\item If there is no risk of deadlock, one proceeds to determine an upper bound. This, as we will see, can be done in \emph{polynomial} time. 

\item Next, one verifies that the upper bound found in the previous step corresponds to an admissible z-chain. If that is the case, the upper bound corresponds to the exact value. This verification procedure can also be carried out in \emph{polynomial} time. 

\item Finally, if the previous steps fail, one needs to search for an admissible resource allocation yielding the maximum blocking time. To that end, one could explore the space of admissible allocations using heuristic-based tree-search, which is a complete method able to compute the blocking time \emph{exactly}, as well as to provide a \emph{proof}, in the format of a z-chain.

\end{enumerate}

%% file: bound.tex
In \cite{Rajkumar}, Rajkumar proposes a branch-and-bound search technique to determine an upper bound $B_i$ on the blocking delay of each job under PIP, assuming that each job can hold at most one resource at a time.
The method consists in summing the durations of the longest critical sections of jobs that can block $J_i$, with the restriction that all jobs must be different and the critical sections must be associated with all different semaphores.

Such an approach has three main limitations: \begin{enumerate} \item it has an \emph{exponential complexity},  \item it \emph{only applies in the absence of nested sections}, and \item it is \emph{not an exact method}, as it only provides an upper bound.\end{enumerate}

In this section we address the first two limitations, by showing how the same upper bound can be computed using a polynomial complexity algorithm, called the Hungarian method \cite{Kuhn,Munkres}, and that such a method does not depend on the number of resources a job can hold at a time. 

\begin{figure}
\begin{algorithmic} [1]

\Function{Blocking-Time-Matrix}{$\Gamma_H,\mathcal{R}_H$}
\State $d \gets$ a new $|\Gamma_H| \times |\mathcal{R}_H|$ zero matrix
\ForAll{$J_j$ \textbf{in} $\Gamma_H$}
	\ForAll{$z_{j,k}$ \textbf{in} $\beta_{j}$ \textbf{such that} $R_{j,k} \in \mathcal{R}_H$}
		\If{$d(J_j,R_{j,k})<d_{j,k}$}
			\State $d(J_j,R_{j,k}) \gets d_{j,k}$
		\EndIf
	\EndFor
\EndFor 
\State \Return $d$
\EndFunction

\bigskip
    
\Function{Cost-Matrix}{$d,\Gamma_H,\mathcal{R}_H$}
\State $N \gets \max \{|\Gamma_H|,|\mathcal{R}_H|\}$
\State $D \gets \max \{ d(J_j,R_k) \}$
\State $m \gets$ a new $N \times N$ matrix
\ForAll{$J_j,R_k$}
	\State $m(J_j,R_k) \gets D - d(J_j,R_k)$
\EndFor
\State \Return $m$
\EndFunction

\bigskip

\Function{H}{$\Gamma_H,\mathcal{R}_H$}
\State    $d \gets \textsc{Blocking-Time-Matrix}(\Gamma_H,\mathcal{R}_H)$
\State    $m \gets \textsc{Cost-Matrix}(d,\Gamma_H,\mathcal{R}_H)$

\Repeat
\State \Comment \textbf{Step 1}: subtract min value $\alpha$ from each row    
\ForAll{$J_j$ \textbf{in} $\Gamma_H$}
    \State $\alpha \gets \min \{ m(J_j,\cdotp) \}$
	\ForAll{$R_k$ \textbf{in} $\mathcal{R}_H$}
		\State $m(J_j,R_k) \gets m(J_j,R_k) - \alpha$
	\EndFor
\EndFor

\State \Comment \textbf{Step 2}: subtract min value $\gamma$ from each column    
\ForAll{$R_k$ \textbf{in} $\mathcal{R}_H$}
	\State $\gamma \gets \min \{ m(\cdotp,R_k) \}$
	\ForAll{$J_j$ \textbf{in} $\Gamma_H$}
		\State $m(J_j,R_k) \gets m(J_j,R_k) - \gamma$
	\EndFor
\EndFor 
\algstore{myalg}
\end{algorithmic}
\caption{Hungarian method (first part)}
\label{alg:hungarian-method}
\end{figure}

\begin{figure}                     
\begin{algorithmic} [1]
\algrestore{myalg}
\State \Comment \textbf{Step 3}: check if assignment is possible    
\State $assignment \gets possible$
\State $\Gamma^* \gets \Gamma_H$
\State $\mathcal{R}^* \gets \mathcal{R}_H$
\State $\mathcal{H} \gets \emptyset$
\State $h \gets 0$
\While{$\Gamma^*\neq\emptyset$ \textbf{and} $assignment$ \textbf{is} $possible$}
	\State $J_j \gets$ job in $\Gamma^*$ \textbf{such that} $m(J_j,\cdotp)$ has $\min$ number of 0 elements
	\If{$\exists R_k\in \mathcal{R}^*$ \textbf{such that} $m(J_j,R_k)=0$}
		\State $\Gamma^* \gets \Gamma^* \setminus \{J_j \}$
		\State $\mathcal{R}^* \gets \mathcal{R}^* \setminus \{R_k \}$
		\State $\mathcal{H} \gets \mathcal{H} \cup \{(J_j,R_k)\}$
		\State $h \gets h + d(J_j,R_k)$
	\Else
		\State $assignment \gets impossible$
	\EndIf
\EndWhile

\State \Comment \textbf{Step 4}: if impossible, transform $m$ and repeat
\If{$assignment$ \textbf{is} $impossible$}
	\State $s \gets \min$  set of rows/cols covering all 0s
    \State $\Theta^- \gets \{ (J_j,R_k)|row(J_j)\notin s, col(R_k) \notin s\}$
    \State $\Theta^+ \gets \{ (J_j,R_k)|row(J_j)\in s, col(R_k) \in s\}$
	\State $\theta \gets \min \{ m(J_j,R_k) | (J_j,R_k) \in \Theta^- \}$
	\ForAll{$(J_j,R_k)$ \textbf{in} $\Theta^-$}
		\State $m(J_j,R_k) \gets m(J_j,R_k) - \theta$
	\EndFor
	\ForAll{$(J_j,R_k)$ \textbf{in} $\Theta^+$}
		\State $m(J_j,R_k) \gets m(J_j,R_k) + \theta$
	\EndFor
\EndIf

\Until{$assignment$ \textbf{is} $possible$}
\State \Return $h$
\EndFunction

\end{algorithmic}
\caption{Hungarian method (continuation)}
\label{alg:hungarian-method-continued}
\end{figure}

The Hungarian method is a combinatorial optimization algorithm that solves the assignment problem  \cite{AssignmentProblem} in polynomial time. Assignment is a minimization problem, described as finding an optimal assignment of \textit{tasks} to \textit{workers}, based on a square \textit{cost} matrix. The problem we address can be considered as an assignment problem's \emph{dual}, where ``tasks'' are resources to be assigned to jobs (the ``workers''), ``costs'' are defined by longest durations, and the objective is to \emph{maximize} (as opposed to minimize, hence the ``dual'' problem) the total time spent by the jobs on these resources.

The method we propose consists in casting the problem into an assignment problem's dual, and then applying the Hungarian method, which we can always do as long as we express the input data in the form of a square cost matrix.

Our algorithm for determining Rajkumar's upper bound using the Hungarian method is shown in Figure~\ref{alg:hungarian-method}. For generality, the algorithm is expressed as a function \textsc{H} with two arguments: a generic set of jobs, $\Gamma_H$, and a generic set of resources, $\mathcal{R}_H$. When invoked with arguments $\Gamma^i$ and $\mathcal{R}^i$, \textsc{H} returns Rajkumar's upper bound for $J_i$'s blocking time, $B_i$. Moreover, when invoked with arguments $\Gamma_N^i$ and $\mathcal{R}_N^i$ it provides an upper bound for $J_i$'s blocking time when each job can hold multiple resources at a time. Finally, in the next section we will see that \textsc{H} also serves a key purpose in the exact  computation of $J_i$'s maximum blocking time, when applied to subsets of $\Gamma^i$ and $\mathcal{R}^i$. Hence our presentation of the algorithm in its parametric format.

Function \textsc{H} relies on two matrices as its main data structures:  a $|\Gamma_H| \times |\mathcal{R}_H|$ \textit{blocking time} matrix, denoted $d$, whose cells contain the longest durations of critical sections, and an $N \times N$ \textit{cost} matrix $m$ constructed from $d$, with $N=\max\{|\Gamma_H|, |\mathcal{R}_H|\}$.

\begin{notation}
For better readability, references to matrix elements are made via their corresponding jobs/resources. So $(J_j,R_k)$ identifies the matrix element corresponding to job $J_j$ and resource $R_k$. The value of such an element in $d$ is denoted by $d(J_j,R_k)$, in $m$ by $m(J_j,R_k)$. Moreover, $col(R_k)$  denotes the index of the column of $m$ corresponding to $R_k$ and  $row(J_j)$ the index of the row of $m$ corresponding to $J_j$.
\end{notation}

In order to obtain a square cost matrix, $m$ is first filled with the opposite of the homologous values in $d$, increased by a constant $D$ to ensure that $m$ only contains positive values, and then padded with 0 rows or columns.

Once $m$ is set up, the Hungarian method is described by the following four steps:

\begin{itemize}
\item[] \textbf{Step 1.} Subtract the smallest element in each row from all the elements of its row. Each row will contain at least one 0 element and no negative element.
\item[] \textbf{Step 2.} Subtract the smallest element in each column from all the elements of its column. Each column will contain at least one 0 element and no negative element.
\item[] \textbf{Step 3.} Check if an assignment is \textit{possible}. An assignment is possible if and only if there is a collection of $N$ 0 elements in $N$ distinct rows and $N$ distinct columns. One way to check  that  is to proceed row by row selecting the row with the least number of 0 elements and mark the (unmarked) column intersecting the first 0 element of that row. If finding a 0 element in each row by only looking at unmarked columns is possible for every row, then the assignment is possible, and the return value, $h$, is computed as the sum of the values of elements in the $d$ matrix corresponding to the 0 elements found in the $m$ matrix. 
\item[] \textbf{Step 4.} If no assignment is possible, transform $m$ and go back to Step 1. To transform $m$, first find a minimum set of rows and columns $s$ that covers all the 0s in $m$. This can be done by applying the method described by Munkres in~\cite{Munkres}, not shown here, in the interest of brevity. Notice that, because no assignment is possible, $|s|<N$. Then, let $\theta$ be the smallest entry in $m$ outside of the rows/columns in $s$. Subtract $\theta$ from each element in $m$ outside of the rows/columns in $s$, and add $\theta$ to each element in $m$ that sits at the intersection of rows/columns in $s$. 
\end{itemize}

The computational complexity of the Hungarian method is  $n^3$, which is much smaller than the $n!$ complexity of the straightforward attack on the problem \cite{Munkres}. Notice that, alongside with computing $h$, \textsc{H} also constructs a set of job/resource pairs $\mathcal{H}$, which will be needed later for check admissibility (see Section~\ref{sec:checking-admissibility}).

\begin{notation}
In the examples that follow, we will use  square brackets to signify relevant sections, with an indication of the associated resource and duration. For instance, with reference to a job $J_j$, the expression $[R_4: 3 ][R_3:2]$ denotes a sequence of two critical sections $\langle z_{j,1}, z_{j,2} \rangle$, where $R_{j,1} = R_4, d_{j,1}=3, R_{j,2}=R_3$, and  $d_{j,2} = 2$.
\end{notation}

\begin{example}
Let us consider a set $\Gamma = \{ J_1, \dots, J_6 \}$, whose jobs access a set of shared resources $\mathcal{R}= \{ R_1, \dots, R_4 \}$, in the following way:
\begin{description}
\item [$J_1$]
$[R_2:1]$
\item [$J_2$]
$[R_4:1] [R_3:1] [R_4:1]$
\item [$J_3$] 
$[R_4:3] [R_3:2]$
\item [$J_4$]
$[R_2:1] [R_1:1] [R_2:1]$
\item [$J_5$]
$[R_3:1] [R_2:1] [R_3:2]$
\item [$J_6$]
$[R_1:2]$
\end{description}
Let us now compute the upper bounds on the blocking times.

For $J_1$, we obtain $\mathcal{R}^1 = \{R_2\}$ and $\Gamma^1 = \{ J_4, J_5 \}$, thus the blocking time matrix $d_1$, with max element 1, and the corresponding cost matrix $m_1$, are as follows:
\[d_1=
\begin{bmatrix}
    \textbf{1} \\
    \textbf{1} \\
\end{bmatrix}
~~~
m_1=
\begin{bmatrix}
    \textbf{0} & 0 \\
    0 & \textbf{0} \\
\end{bmatrix}
\]
Since $m_1'$ contains two 0s in two distinct rows/columns, we obtain $B_1=d_1(J_4,R_2)=1$.

For $J_2$, we obtain $\mathcal{R}^2 = \{R_2, R_3, R_4\}$ and $\Gamma^2 = \{ J_3, J_4, J_5 \}$, thus matrix $d_2$, with max element 3, and corresponding cost matrix before ($m_2$) and after Step 1 ($m_2'$) are as follows:
\[d_2=
\begin{bmatrix}
    0 & 2 & \textbf{3} \\
    1 & 0 & 0 \\
    1 & 2 & 0 \\
\end{bmatrix}
~~~
m_2=
\begin{bmatrix}
    3 & 1 & 0 \\
    2 & 3 & 3 \\
    2 & 1 & 3 \\
\end{bmatrix}
~~~
m_2'=
\begin{bmatrix}
    3 & 1 & \textbf{0} \\
    \textbf{0} & 1 & 1 \\
    1 & \textbf{0} & 2 \\
\end{bmatrix}
\]

Since $m_2'$ contains three 0s in three distinct rows/columns, at positions $m_2'(J_3,R_4)$, $m_2'(J_4,R_2)$ and $m_2'(J_5,R_3)$ as indicated in bold, we can conclude that $B_1=d_2(J_3,R_4)+d_2(J_4,R_2)+d_2(J_5,R_3)=6$.

With the other jobs we obtain:
$\mathcal{R}^3 = \{R_2, R_3\}, \Gamma^3 = \{ J_4, J_5 \}$
\[d_3=
\begin{bmatrix}
    1 & 0 \\
    1 & \textbf{2} \\
\end{bmatrix}
~~~
m_3=
\begin{bmatrix}
    1 & 2 \\
    1 & 0 \\
\end{bmatrix}
~~~
m_3'=
\begin{bmatrix}
    \textbf{0} & 1 \\
    1 & \textbf{0} \\
\end{bmatrix}
\]

Thus $B_3=d_3(J_4,R_2)+d_3(J_5,R_3)=3$.
$\mathcal{R}^4 = \{R_1, R_2, R_3\}, \Gamma^4 = \{ J_5, J_6 \}$
\[d_4=
\begin{bmatrix}
    0 & 1 & \textbf{2} \\
    2 & 0 & 0 \\
\end{bmatrix}
~~~
m_4=
\begin{bmatrix}
    2 & 1 & \textbf{0} \\
    \textbf{0} & 2 & 2 \\
    0 & \textbf{0} & 0 \\
\end{bmatrix}
\]

Thus $B_4=d_4(J_5,R_3)+d_4(J_6,R_1)=4$.
Finally,
$\mathcal{R}^5 = \{R_1\}, \Gamma^5 = \{ J_6 \}, B_5=d(J_6,R_1)=2$ and
$B_6=0$ (trivially).
\label{ex:faldella-Asix}
\end{example}

The upper bounds found in Example~\ref{ex:faldella-Asix} coincide with maximum blocking times. However, in general the blocking time could be less than the upper bound, as shown by the following example.

\begin{example}
Let $\Gamma$ be $\{ J_1, \dots, J_4 \}$ and let its jobs access  resources in $\mathcal{R}= \{ R_1, R_2 \}$, in the following way:
\begin{description}
\item [$J_1$]
$[R_1:4] [R_2:5]$
\item [$J_2$]
$[R_2:4] [R_1:3]$
\item [$J_3$] 
$[R_1:1] [R_2:3]$
\item [$J_4$]
$[R_2:1]$
\end{description}
For $J_1$ we obtain $\mathcal{R}^1 = \{R_1, R_2\}$ and $\Gamma^2 = \{ J_2, J_3, J_4 \}$, thus matrix $d$, with max element $d(J_2,R_2)=4$, and corresponding cost matrix before Step 1 ($m$) and after Step 2 ($m'$) are as follows:
\[d=
\begin{bmatrix}
    3 & \textbf{4} \\
    1 & 3  \\
    0 & 1  \\
\end{bmatrix}
~~~
m=
\begin{bmatrix}
    1 & 0 & 0 \\
    3 & 1 & 0 \\
    4 & 3 & 0 \\
\end{bmatrix}
~~~
m'=
\begin{bmatrix}
    0 & 0 & \underline{0} \\
    2 & \textbf{1} & 0 \\
    3 & 3 & 0 \\
\end{bmatrix}
\]
Since $m'$ does not contains three 0s in three distinct rows/columns, the first row and the last column alone cover all of its 0s. The smallest value in the remaining cells is 1 (indicated in bold). Therefore after Step 3 we obtain $m''$ by adding 1 to the cell at the intersection of the covering rows/columns (underlined), and subtracting 1 from the cells outside of the covering rows/columns.
\[m''=
\begin{bmatrix}
    \textbf{0} & 0 & 1 \\
    1 & \textbf{0} & 0 \\
    2 & 2 & \textbf{0} \\
\end{bmatrix}
\]

Because $m''$ contains three 0s in different rows/columns, as indicated in bold, we determine $B_1=d(J_2,R_1)+d(J_3,R_2)=3+3=6$. 

However, this upper bound refers to an impossible resource allocation, one associated with an inadmissible z-chain $\langle z_{2,2}, z_{3,2} \rangle$ (while $J_3$ holds $R_2$, $z_{2,1}$ obstructs $z_{2,2}$).
\label{ex:faldella-II}
\end{example}

If each job can hold multiple resources at a time, the same method can be used to compute a bound by simply using $\Gamma_N^i$ and $\mathcal{R}_N^i$ as parameters.

\begin{example}
Let us consider an application $\Gamma = \{ J_1, \dots, J_6 \}$. Let jobs in $\Gamma$ access a set of shared resources $\mathcal{R}= \{ R_1, \dots, R_4 \}$, in the following way:
\begin{description}
\item [$J_1$]
$[R_2:1]$
\item [$J_2$]
$[R_4:3 ~ [R_3:1]]$
\item [$J_3$] 
$[R_4:3] [R_3:2]$
\item [$J_4$]
$[R_2:3 ~ [R_1:1]]$
\item [$J_5$]
$[R_3:4 ~ [R_2:1]]$
\item [$J_6$]
$[R_1:2]$
\end{description}
We observe that $z_{2,2} \subset z_{2,1}$, $z_{4,2} \subset z_{4,1}$, and $z_{5,2} \subset z_{5,1}$, which is compatible with resource ordering $R_1 < R_2 < R_3 < R_4$.
Let us now focus on $B_2$. We observe that $\mathcal{R}^2 = \{R_2, R_3, R_4\}$, $\mathcal{R}_N^2 = \{R_1, R_2, R_3, R_4\}$, $\Gamma^2 = \{ J_3, J_4, J_5 \}$, and $\Gamma_N^2 = \{ J_3, J_4, J_5, J_6 \}$. We obtain the following blocking time matrix:
\[d=
\begin{bmatrix}
    0 & 0 & 2 & \textbf{3} \\
    1 & \textbf{3} & 0 & 0 \\
    0 & 1 & \textbf{4} & 0 \\
    \textbf{2} & 0 & 0 & 0 \\
\end{bmatrix}
\]
whereupon we can easily identify $B_2=3+3+4+2=12$.
\label{ex:faldella-Asixprime}
\end{example}

Notice that when each job can hold multiple resources at a time, using a bound such as this one could  lead to a significant overestimation of the blocking time, since we are considering some resources, in particular those belonging to $\mathcal{R}_N^i \setminus \mathcal{R}^i$, whose potential for causing blocking really depends on the whole z-chain. This can be seen in Example~\ref{ex:faldella-Asixprime}, where  $R_1 \in \mathcal{R}_N^i \setminus \mathcal{R}^i$ only appears, within $\Gamma^2_N$, in $J_4$, and it does so in a section contained by $z_{4,1}$. Therefore, if $J_4$ hasn't entered $z_{4,1}$ before $J_2$ is activated, $J_2$ cannot be blocked because of $R_1$.

%% file: admissibility.tex
To check that the bound found by the Hungarian method is matched by a possible resource allocation that can block $J_i$, we can start from the set $\mathcal{H}$ produced by \textsc{H}, in order to construct a z-chain $\mathcal{Z}$ corresponding to the selection of  cells in the blocking time matrix that yields the bound. We do so incrementally, by making sure that the so-constructed $\mathcal{Z}$ is induction compatible. Once we have $\mathcal{Z}$, we shall scan it by ascending priority in order to check that each element satisfies FLO. If that is the case, $\mathcal{Z}$ is admissible, thus the bound corresponds to a possible resource allocation that can block $J_i$, and the blocking time matches the bound. If, however, we cannot construct an admissible $\mathcal{Z}$, the admissibility check fails. By following this procedure, we can prove that the bound obtained for Example~\ref{ex:faldella-Asixprime} corresponds to an admissible z-chain, therefore $B_2=12$ is the actual blocking time of $J_2$. The \textsc{Admissible} function defined in Figure~\ref{alg:admissibility-check-nested-sections} implements a polynomial-time, heuristic procedure for checking admissibility. The $\mathcal{Z}$ produced by it satisfies by construction NBJ, NBR (because $z \in \mathcal{Z}$ correspond to elements in $\mathcal{H}$ associated with all-different resources and jobs) and induction compatibility (because of the $R \in \mathcal{R}_{\mathcal{I}}$ condition). Moreover, lines \ref{admissible:from}-\ref{admissible:to} ensure that FLO, FHO and LSM also hold. It should be noticed, however, that such a procedure is sound but not complete, as the following example shows.

\begin{example}
Let us consider an application $\Gamma = \{ J_1, \dots, J_6 \}$. Let jobs in $\Gamma$ access a set of shared resources $\mathcal{R}= \{ R_1, \dots, R_4 \}$, in the following way:
\begin{description}
\item [$J_1$]
$[R_2:1]$
\item [$J_2$]
$[R_2:2] ~ [R_2:2 ~ [R_1:1]]$
\item [$J_3$] 
$[R_1:2]$
\end{description}
We observe that $z_{2,3} \subset z_{2,2}$  which is compatible with resource ordering $R_1 < R_2$.
Let us now focus on $B_1$. We observe that $\mathcal{R}_N^1 = \{R_1, R_2\}$ and $\Gamma_N^1 = \{ J_2, J_3 \}$. We obtain the following blocking time matrix:
\[d=
\begin{bmatrix}
    1 & \textbf{2} \\
    \textbf{2} & 0 \\
\end{bmatrix}
\]
with  $\mathcal{H} = \{(J_2,R_2), (J_3,R_1)\}$ and $h=4$. Such a value corresponds indeed to the exact value of $B_1$, since there exists an admissible $\mathcal{Z} = \langle z_{2,2}, z_{3,1}  \rangle$  where the allocation of resources to jobs is $R_2/J_2, R_1/J_3$ and $d(\mathcal{Z})=4$. However, the admissibility check fails, because \textsc{Admissible} only finds the left-most section  $z_{2,k}$ with duration $d_{2,k}=d(J_2,R_2)=2$, which is $z_{2,1}$, and does not consider other options. However, given $\langle z_{2,1}, z_{3,1}  \rangle$, $z_{3,1}$ is not induction compatible, thus $\langle z_{2,1}, z_{3,1}  \rangle$ is not admissible.
\label{ex:admissibility-incompleteness}
\end{example}

In general, a complete admissibility check with nested sections may require several backtracks, which  increase its complexity and thus lose its purpose. It seems therefore more effective to try a simple heuristic method first and then, if that fails, proceed with the exact method we will present next.
Also notice that \textsc{Admissible} is complete if any of the following conditions holds:
\begin{itemize} 
\item $\mathcal{R}^i= \mathcal{R}_N^i$ (as it is the case in the absence of nested sections); or
\item $\forall (J_j,R) \in \mathcal{H}$, there is only one $z_{j,p}\in \beta_j$ such that $R_{j,p}=R$ and $d_{j,p}=d(J_j,R)$.
\end{itemize}

\begin{figure}
\begin{algorithmic} [1]
\Function{Admissible}{$\Gamma,d,\mathcal{H},h$} 
\State \Comment Determine if $\mathcal{H}$ corresponds to an admissible z-chain
	\State $\mathcal{Z} \gets \emptyset$
    \State $d \gets 0$
    \State $\mathcal{R}_\mathcal{I} \gets \mathcal{R}^i$ 
    \While {$\exists (J_j,R)\in\mathcal{H}$ \textbf{such that} $R \in \mathcal{R}_\mathcal{I}$ \textbf{and} $d(J_j,R)>0$}
    \State $k \gets 1$ \label{lst:line:blah2}
    \State \textit{found} $\gets$ \textit{false}
    \While{\textit{found} \textbf{is} \textit{false}}
    \If{$R_{j,k}=R$ \textbf{and} $d_{j,k}=d(J_j,R)$}
    \State \textit{found} $\gets$ \textit{true}
    \State $d \gets d + d(J_j,R)$
    \State  $\mathcal{Z} \gets \mathcal{Z} + \langle z_{j,k} \rangle$
    \State $\mathcal{R}_\mathcal{I} \gets \mathcal{R}_\mathcal{I} \cup \{ R_{j,q} | z_{j,q} \subset z_{j,k} \} \setminus \{ R \}$
    \Else
    \State $k \gets k+1$
    \EndIf
    \EndWhile
    \EndWhile
    \If{$d<h$}
    \State \Return \textit{false}
    \EndIf
    \State $j \gets |\Gamma|$ \label{admissible:from}
    \State $\mathcal{R}_\mathcal{Z} \gets \emptyset$
    \While{$j>i$}
    \If{$\exists z_{j,p} \in \mathcal{Z}$}
    \For{$q=1$ \textbf{to} $p$}
    \If{$R_{j,q} \in \mathcal{R}_\mathcal{Z}$}
    \State \Return \textit{false} \Comment Violation of FLO
    \EndIf
    \EndFor
    \State $\mathcal{R}_\mathcal{Z} \gets \mathcal{R}_\mathcal{Z} \cup \{  R_{j,s} | z_{j,s} \supseteq z_{j,p}  \}$
    \EndIf
    \State $j \gets j-1$ 
    \EndWhile \label{admissible:to}
    \State \Return \textit{true}
\EndFunction

\end{algorithmic}
\caption{Admissibility check for nested sections}
\label{alg:admissibility-check-nested-sections}
\end{figure}

%% file: computation.tex
To compute $J_i$'s maximum blocking time $B_i$ we can apply $A^*$ \cite{AStar}, which is a heuristic-based, exact search algorithm \cite{RussellNorvig}. $A^*$ is defined in general for graphs. However, we can gain in simplicity and efficiency by exploiting the tree structure of the search space resulting from the absence of nested sections. 

The data structure used by $A^*$ is a \textit{search tree}, where each $node$ is associated with an admissible z-chain that uniquely defines a (partial) \textit{allocation of resources} to jobs. Nodes can be extended by extending the z-chain, leading to more nodes.
The \textit{root} of the search tree is the \textit{empty node}, where no resources are allocated. 
The \textit{gain} $g$ of a $node$ is equal to the duration of the z-chain.
Terminal (\emph{leaf}) nodes  are those associated with z-chains that have no admissible extensions.
An \emph{optimal solution} corresponds to a node associated with a z-chain with the longest duration. Only terminal nodes represent optimal solutions. The likelihood of a node to lead to an optimal solution is estimated by the duration of the z-chain (gain) plus the estimated duration of its longest extension (\emph{heuristic value}), considering the remaining jobs and resources.

A key aspect of $A^*$ is that the search tree is not all generated blindly at start, because that would mean creating and keeping in memory an exponentially large number of nodes. Instead, only one node is expanded at a time. The node is selected among a set of candidate nodes for expansion, called \emph{fringe}, according to the estimated gain. 

\begin{definition}[Estimated gain]
The \textbf{estimated gain} of a solution through a given $node$ is $f(node)$ = $node.g$ + $node.h$.  
\end{definition}

\begin{definition}[Fringe]
The \textbf{fringe} \cite{RussellNorvig} is the ordered collection of nodes that have been generated but not yet expanded. 
\end{definition}

\begin{definition}[Node]
A $node$ in the search tree is a data structure $$\langle \mathcal{Z}, \mathcal{R}_\mathcal{Z}, \Gamma_\mathcal{Z}, \mathcal{R}_H, \Gamma_H, \mathcal{R}^{in}, \Gamma^{in}, g, h \rangle$$ where: \begin{itemize}
\item [$\mathcal{Z}$] is the set of critical sections that have been explored so far in the current branch of the search tree, encoding an allocation of resources in $\mathcal{R}_N^i$ to jobs in $\Gamma_N^i$; 
\item [$\mathcal{R}_\mathcal{Z}$] is the set of resources associated with sections in $\mathcal{Z}$; 
\item [$\Gamma_\mathcal{Z}$] is the set of jobs associated with sections in $\mathcal{Z}$;\footnote{$\mathcal{R}_\mathcal{Z}$, $\Gamma_\mathcal{Z}$, as well as other fields of the $node$ structure, could be derived from $\mathcal{Z}$, but are kept separate for efficiency.} 
\item [$\mathcal{R}_H$] is  $\mathcal{R}_N^i \setminus \mathcal{R}_\mathcal{Z}$;
\item [$\Gamma_H$] is $\Gamma_N^i \setminus \Gamma_\mathcal{Z}$;
\item [$\mathcal{R}^{in}$] is $in(J_i,\mathcal{Z})$; 
\item [$\Gamma^{in}$] is the set of jobs in $\Gamma_H$ containing sections that are maximal with respect to $\mathcal{R}^i$ and are associated with  resources that do not belong to $\mathcal{R}_\mathcal{Z}$;
\item [$g$] is the total duration of all sections in $\mathcal{Z}$ (gain); 
\item [$h$] is the heuristic value associated with  $node$.
\end{itemize} 
\label{def:node}
\end{definition}

\begin{figure}[t]
\centering
\includegraphics[width=.45\columnwidth]{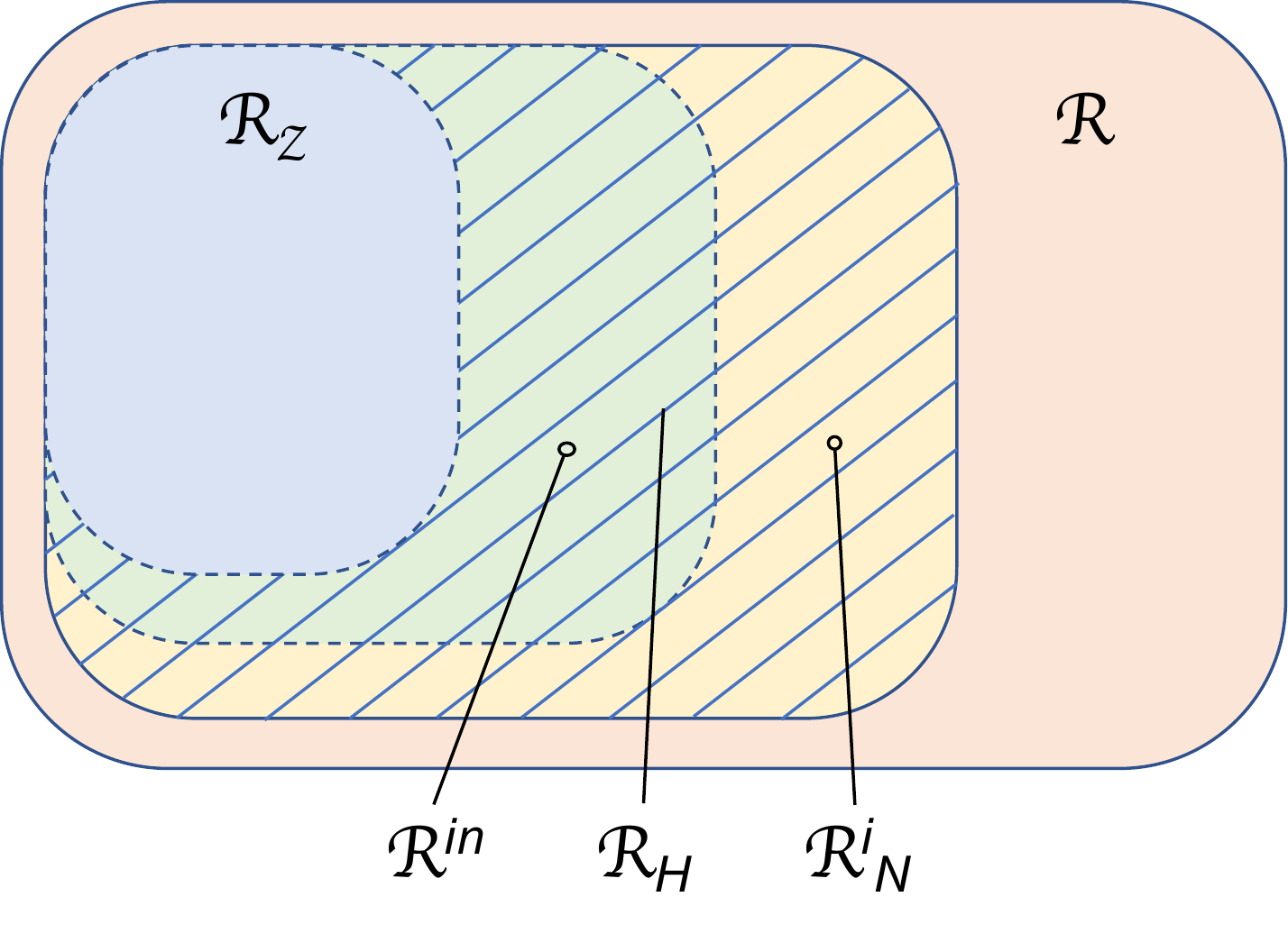}
\caption{Relations among resource sets.}
\label{fig:resource-sets}
\end{figure}

The relations among sets of resources relevant to Definition \ref{def:node} are illustrated in Figure~\ref{fig:resource-sets}.
The \emph{heuristic value} associated with a $node$ is the bound on the maximum blocking time produced by jobs in $node.\Gamma_H$ via resources in $node.\mathcal{R}_H$,\footnote{Henceforth, we will use  the dot notation to identify elements of a node: $node.\mathcal{Z}$, $node.\mathcal{R}_\mathcal{Z}$, etc.} and it can be determined in polynomial time using the method seen in Section~\ref{sec:bound}. Notice that $node.h$ is 0 if and only if $node$ is a leaf node.

\begin{figure}[t]
\begin{algorithmic} [1]
\Function{Blocking-Time}{$\Gamma,i$} 
	\State $\mathcal{R}_N^i \gets$ \textsc{Relevant-Resources}$(\Gamma,i)$
    \State $\Gamma_N^i \gets \{ J_j | j>i \textrm{ and } J_j  \textrm{ uses resources in } \mathcal{R}_N^i \}$
    \State $h_0 \gets \textsc{H}(\Gamma_N^i,\mathcal{R}_N^i)$
    \State $n_0 \gets \langle
	    \emptyset, 
	    \emptyset, 
	    \emptyset, 
	    \mathcal{R}_N^i, 
	    \Gamma_N^i, 
	    \mathcal{R}^i, 
	    \Gamma^i, 
	    0, 
	    h_0
    \rangle$
    \State $\textit{fringe} \gets \langle n_0 \rangle$     
\While{\emph{true}} 
       	\State $n \gets$ \textsc{Remove-First}(\textit{fringe}) 
        	\If{$n.h = 0$}
           	\State 
            \Return $n.g$
        \Else\
        \textsc{Insert-All}(\textsc{Expand}($n$),\textit{fringe})
        \EndIf
    \EndWhile
\EndFunction
\end{algorithmic}
\caption{Blocking time computation: main function}
\label{alg:blocking-time}
\end{figure}

\begin{figure}[t]
\begin{algorithmic} [1]
\Function{Expand}{$n$}
	\State \textit{successors} $ \gets \emptyset$
    \ForAll{$z_{j,p}$ \textbf{in} \textsc{Successors}($n$)}
    	\State $s \gets$ a new node
        \State $s.\mathcal{Z} \gets n.\mathcal{Z} \cup \{z_{j,p}\}$
		\State $s.\mathcal{R}_\mathcal{Z} \gets n.\mathcal{R}_\mathcal{Z} \cup \{R_{j,p}\}$
          \State $s.\Gamma_\mathcal{Z} \gets  n.\Gamma_\mathcal{Z} \cup \{J_j\}$
		\State $s.\mathcal{R}_H \gets n.\mathcal{R}_H \setminus \{R_{j,p}\}$
        \State $s.\Gamma_H \gets n.\Gamma_H \setminus \{J_{j}\}$
		\State $s.\mathcal{R}^{in} \gets n.\mathcal{R}^{in} \cup in(J_i,z_{j,p}, n.\mathcal{R}^{in})$
        \State $s.\Gamma^{in} \gets \{ J_k \in s.\Gamma_H |  \beta_k(s.\mathcal{R}^{in}) \setminus \beta_k(n.\mathcal{R}_\mathcal{Z}) \neq \emptyset \}$
        \State $s.g \gets n.g + d_{j,p}$
        \If{$s.\Gamma^{in} = \emptyset$}
        \State 
        $s.h \gets 0$
        \Else\
		\State 
        $s.h \gets$ \textsc{H}$(s.\Gamma_H,s.\mathcal{R}_H)$
        \EndIf
    	\State add $s$ to \textit{successors} 
		\If{$s.h=0$ \textbf{and} $f(s)=f(n)$}
        \State 
        \Return \textit{successors}
        \EndIf
    \EndFor
    \If {\textit{successors} $ =\emptyset$}
    \State $n.h \gets 0$
    \State add $n$ to \textit{successors}
    \EndIf
    \State \Return \textit{successors} 
\EndFunction
\end{algorithmic}
\caption{Blocking time computation: node expansion}
\label{alg:blocking-time-expand}
\end{figure}

\begin{figure}                     
\begin{algorithmic} [1]
\Function{Successors}{$n$}
	\State \Comment Determine all admissible extensions to $n$
    \State \textit{extensions} $ \gets \emptyset$
    \ForAll{$J_j$ \textbf{in} $n.\Gamma^{in}$} \Comment NBJ
	    \ForAll{$z_{j,p}$ \textbf{in} $\beta_{j}(n.\mathcal{R}^{in})\setminus \beta_{j}(n.\mathcal{R}_\mathcal{Z})$} \Comment NBR, LSM
        \If{$\exists \hat{n} \in \textit{fringe}$ \textbf{such that} $n.\mathcal{Z} + \langle z_{j,p} \rangle \subseteq\hat{n}.\mathcal{Z}$} \label{algo:successors:here}
        \State \Comment Avoid considering the same z-chain twice 
        \State discard $z_{j,p}$ and continue 
            \EndIf
        \State $\mathcal{R}^s \gets \{ R_{j,s} | z_{j,s} \supseteq z_{j,p} \}$
        \ForAll{$z_{h,q} | z_{h,r} \in n.\mathcal{Z}, h<j, q<r$}
        \If{$R_{h,q} \in \mathcal{R}^s$}
        \State discard $z_{j,p}$ and continue \Comment FHO
        \EndIf
        \EndFor
        \ForAll{$J_l \in n.\Gamma_{\mathcal{Z}}$ \textbf{such that} $j<l$}
        \State $\mathcal{R}^q \gets \{ R_{l,q} | z_{l,r} \in n.\mathcal{Z}, z_{l,q} \supseteq z_{l,r} \}$
        \ForAll{$o<p$}
        \If{$R_{j,o}\in \mathcal{R}^q$}
        \State discard $z_{j,p}$ and continue \Comment FLO
        \EndIf
        \EndFor
        \EndFor
    		\State add $z_{j,p}$ to \textit{extensions} 
	    \EndFor
    \EndFor
	\State \Return \textit{extensions} 
\EndFunction
\end{algorithmic}
\caption{Blocking time computation: identification of admissible extensions}
\label{alg:blocking-time-successors}
\end{figure}

The exact algorithm for computing the maximum blocking time $B_i$ of a job $J_i$ is shown in Figure~\ref{alg:blocking-time}.
Initially, the fringe only contains the empty node 
$\langle 
	    \emptyset, 
	    \emptyset, 
	    \emptyset, 
	    \mathcal{R}_N^i, 
	    \Gamma_N^i, 
	    \mathcal{R}^i, 
	    \Gamma^i, 
	    0, 
	    h_0
    \rangle$, 
with $h_0$ equal to the upper bound obtained using the method seen in Section~\ref{sec:bound}. The fringe gets populated by new nodes until an optimal solution is reached. To ensure an optimally efficient exploration of the search tree, the elements in the fringe must be kept ordered by descending $f$. To this end, two functions are defined to manipulate \textit{fringe} elements:
\begin{itemize}
    \item \textsc{Remove-First}(\textit{fringe}), which returns the first node in the fringe and at the same time removes it from the fringe;
	\item \textsc{Insert-All}(\textit{nodes},\textit{fringe}), which inserts in the fringe a set of nodes, ensuring that the fringe is kept ordered by descending $f$, and resolving ties arbitrarily but always primarily in favor of leaf nodes, and, secondarily, in favor of newest nodes.\footnote{In order to optimize memory usage, the fringe could be set to contain at most one leaf node, and nothing after that. In this way, each time a new node $n$ is produced, if the last node of the fringe is a leaf node with an estimated gain that exceeds or equals $f(n)$, then $n$ can be simply discarded; else if $n$ is a leaf node then all nodes with an estimated gain no larger than $f(n)$ are removed from the fringe and $n$ is appended to the fringe, becoming its last element; else $n$ is inserted in the fringe before any other node with a lower or equal value of $f$. This technique implements a simplified form of \textit{memory-boundedness} \cite{RussellNorvig}. However, we will keep in the fringe structure \emph{all} the generated and yet unexpanded nodes, in order to be able to ensure that the same z-chain is not explored twice (see function \textsc{Successors}, line \ref{algo:successors:here}).}
\end{itemize}

\textsc{Expand} (Figure~\ref{alg:blocking-time-expand}) creates a set of (non-leaf) \textit{successor} nodes corresponding to the admissible extension found by  \textsc{Successors} (Figure~\ref{alg:blocking-time-successors}).
 
If \textsc{Successors}($node$) $ =\emptyset$, then $node$ is marked as a leaf node: $node.h \gets 0$, and reinserted in the fringe. Otherwise, $node$ is removed from the fringe and the set of successor nodes created by \textsc{Expand}($node$) is added to the fringe.

The algorithm terminates when the element removed from the fringe is a leaf node, whereby the maximum blocking time is returned as that node's gain.

\begin{remark}
Termination of the methods used is proven in \cite{AStar,Munkres}.
\end{remark}

Before we illustrate the method with an example, it is worthwhile  commenting on the method's optimality.

\begin{remark}
Since $h$ is defined as an upper bound, it \textit{never underestimates} the blocking time, and therefore it is an \textbf{admissible heuristic} according to Hart et al.~\cite{AStar}. Because $h$ is admissible, the tree-search $A^*$ method is \textbf{provably optimal}, thus it returns the maximum blocking time (not simply a bound), as well as \textbf{optimally efficient}, thus no other optimal algorithm that uses $h$ as a heuristic is guaranteed to expand fewer nodes \cite{AStar}.
\end{remark}

To illustrate the procedure, let us consider the following example.

\begin{example}
Let us consider an application $\Gamma = \{ J_1, \dots J_5 \}$. Let jobs in $\Gamma$ access a set of shared resources $\mathcal{R} = \{ R_1,\dots, R_5 \}$, in the following way:
\begin{description}
\item[$J_1$] $[R_4:1]$
\item[$J_2$] $[R_4:6 ~[R_3: 4 ~[R_2:2]]]$
\item[$J_3$] $[R_1:5] ~[R_5: 13 ~[R_4:10]]$
\item[$J_4$] $[R_3:3 ~[R_1:1]] ~[R_5:1] ~[R_4:12 ~[R_2:9]]$
\item[$J_5$] $[R_1:4] ~[R_5:13 ~[R_2: 12]] ~[R_1:7]$
\end{description}
We observe that nesting of critical sections is compatible with resource ordering $R_1 < R_2 < R_3 < R_4 < R_5$, and that 
$\mathcal{R}^1=\{R_4\}$,  
$\Gamma^1 = \{ J_2, J_3, J_4 \}$, 
$\mathcal{R}_N^1=\{R_1, R_2, R_3, R_4\}$, and 
$\Gamma_N^1=\{ J_2, J_3, J_4, J_5 \}$.

Using the Hungarian method, we compute $h_0=33$ corresponding to $\mathcal{H} = \{(J_2, R_3), (J_3, R_1),$ $(J_4, R_4), (J_5, R_2) \}$. However, the corresponding z-chain is not admissible, as can be found out by running the \textsc{Admissible} procedure on the data. 

\input{search-tree.tex}

In order to compute the blocking time we therefore construct an initial node $n_0 = \langle \emptyset, \emptyset, \emptyset,$ $\{R_1, R_2, R_3, R_4\}, \{ J_2, J_3, J_4, J_5 \}, \{R_4\}, \{ J_2, J_3, J_4 \}, 0, 33  \rangle$ which constitutes the only element in the starting \emph{fringe}, $\langle n_0 \rangle$.

Node $n_0$ has three possible extensions: $z_{2,1}$, $z_{3,3}$, and $z_{4,4}$. Since $n_0.\mathcal{Z}=\emptyset$, the initial expansions automatically satisfy conditions NBR, LSM, FHO, and FLO, which don't need further checking.

Three new nodes are created accordingly:
\begin{itemize}
\item $n_1 = \langle \{z_{2,1}\}, \{R_4\}, \{J_2\}, \{R_1, R_2, R_3 \}, \{ J_3, J_4, J_5 \}, \{R_2, R_3, R_4\}, \{ J_4, J_5 \}, 6, 20 \rangle$;
\item $n_2 = \langle  \{z_{3,3}\}, \{R_4\}, \{J_3\}, \{R_1, R_2, R_3 \}, \{ J_4, J_5 \}, \{ R_4 \}, \emptyset, 10, 0  \rangle$;
\item $n_3 = \langle  \{z_{4,4}\}, \{R_4\}, \{J_4\}, \{R_1, R_2, R_3 \}, \{ J_2, J_3, J_5 \}, \{R_2, R_4\}, \{ J_5 \}, 12, 21 \rangle$.
\end{itemize}

We have $f(n_1)=26$, $f(n_2)=10$, and $f(n_3)=33$, therefore $fringe=\langle n_3, n_1, n_2 (\star) \rangle$.\footnote{For convenience, $(\star)$ marks the first leaf node in the fringe.}
The first node in the fringe is $n_3$ and is not a leaf node. Its only possible extension satisfying NBJ and NBR is $z_{5,3}$, which is also limited-scope maximal (LSM). However, since $R_{4,3}=R_{5,2}$, $z_{5,3}$ violates FHO. Since $n_3$ has no admissible extensions, we set $n_3.h$ to $0$, obtaining $f(n_3) = 12$ with $n_3$ a leaf node. The $fringe$ becomes $\langle  n_1, n_3 (\star), n_2 \rangle$.

The first node in the fringe is $n_1$ and is not a leaf node. Its possible LSM extensions satisfying NBJ and NBR are $z_{4,1}$  and $z_{5,3}$, which satisfy FHO, because $n_1.\mathcal{Z}$ only contains one section which is not preceded by any other section, as well as FLO, because the only job in $n_1.\Gamma_{\mathcal{Z}}$ is $J_2$, which has higher priority than $J_4$ and $J_5$. Therefore, $z_{4,1}$ and $z_{5,3}$ are both admissible extensions. 
Two new nodes are created accordingly:

\begin{itemize}
\item $n_4 = \langle \{z_{2,1}, z_{4,1}\}, \{R_3, R_4\}, \{J_2, J_4\}, \{R_1, R_2 \}, \{ J_3, J_5 \}, \{R_1, R_2, R_3, R_4\}, \{ J_3, J_5 \}, 9, 17 \rangle$;
\item $n_5 = \langle \{z_{2,1}, z_{5,3}\}, \{R_2, R_4\}, \{J_2, J_5\}, \{R_1, R_3 \}, \{ J_3, J_4 \}, \{R_2, R_3, R_4 \}, \{ J_4 \}, 18, 8 \rangle$,
\end{itemize}

with $f(n_4)=f(n_5)=26$.

We now have $fringe=\langle n_4, n_5, n_3 (\star), n_2 \rangle$. The first node in the fringe is $n_4$ and is not a leaf node. It has 4 possible admissible extensions: $z_{3,1}$, $z_{5,1}$, $z_{5,3}$, and $z_{5,4}$, corresponding  to 4 new nodes: $n_6, \dots, n_9$. Two of these nodes are leaf nodes: $n_7$, with $f(n_7)=13$, and $n_9$, with $f(n_9)=16$, whereas the other two nodes are expandable, with $f(n_6)=f(n_8)=26$. The new fringe is therefore $\langle n_6, n_8, n_5, n_9 (\star), n_7, n_3, n_2 \rangle$, with $n_6.\mathcal{Z}=\langle z_{2,1}, z_{4,1}, z_{3,1} \rangle$. We have only one admissible extension to $n_6$, which is $z_{5,3}$, thus obtaining a last (leaf) node, $n_{10}$, whose associated z-chain  is $n_{10}.\mathcal{Z}=\langle z_{2,1}, z_{4,1}, z_{3,1}, z_{5,3} \rangle$, yielding for $J_1$ a blocking time $B_1=n_{10}.g=d(n_{10}.\mathcal{Z})=26$.

We shall notice how, in order to find the maximum blocking time, we had to explore 11 nodes, as shown in Figure~\ref{fig:search-tree}, whereas an uninformed search of the space would mean evaluating $\prod_{j>1} ( |\beta_j|+1 )=4\times4\times6\times5=480$ possibilities.
\label{ex:last-nested-sections}
\end{example}

%% file: search-tree.tex
\begin{figure}[t]
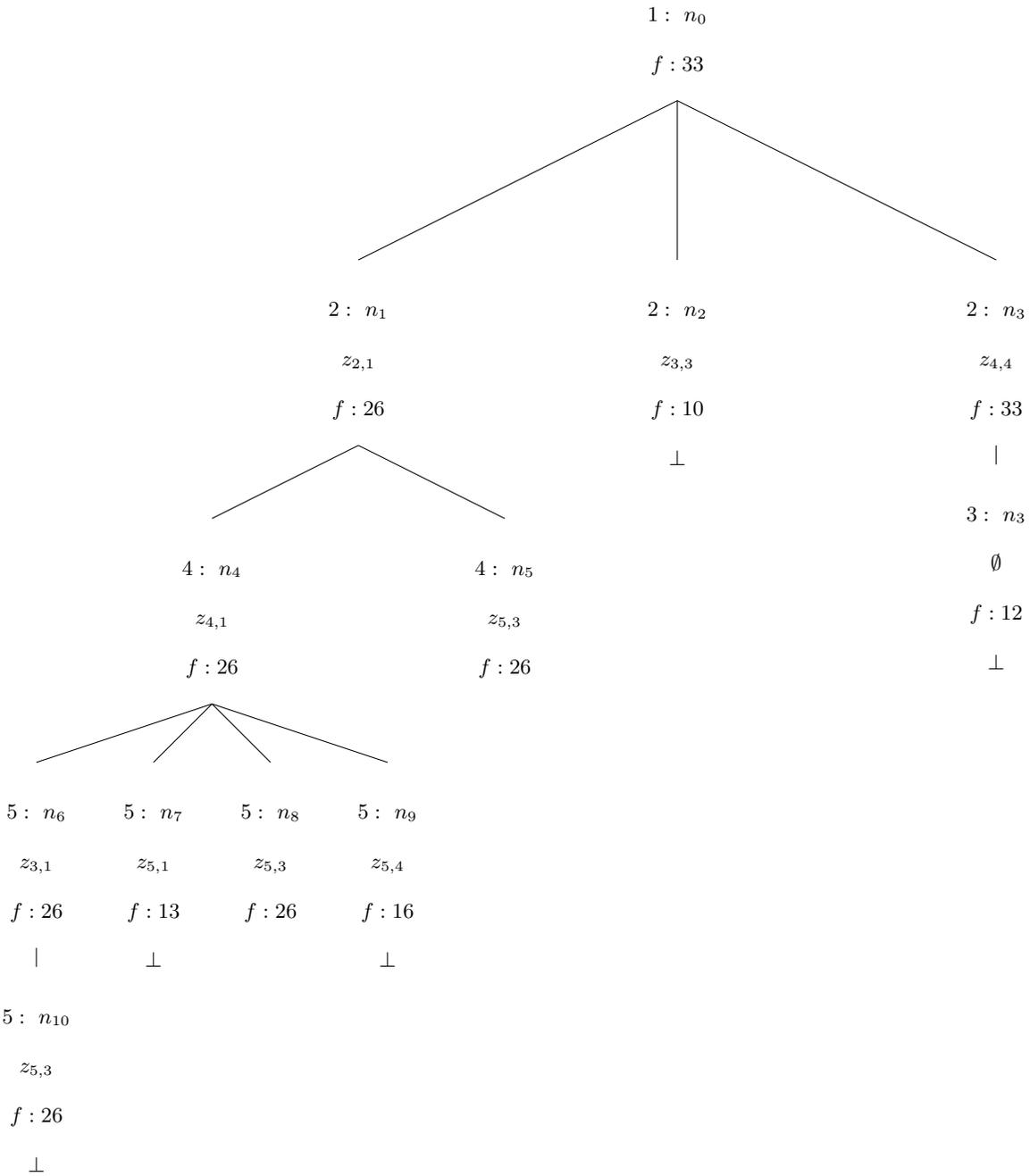

\begin{footnotesize}
\Tree
[.\ensuremath{\parbox{.08\columnwidth}{\begin{gather*}
1:~n_0 \\
f:33
\end{gather*}}}
[.\ensuremath{\parbox{.08\columnwidth}{\begin{gather*}
2:~n_1  \\
z_{2,1} \\
f:26
\end{gather*}}}
[.\ensuremath{\parbox{.08\columnwidth}{\begin{gather*}
4:~n_4 \\
z_{4,1} \\
f:26
\end{gather*}}}
[.\ensuremath{\parbox{.08\columnwidth}{\begin{gather*}
5:~n_6 \\
z_{3,1} \\
f:26
\end{gather*}}}
[.\ensuremath{\parbox{.08\columnwidth}{\begin{gather*}
5:~n_{10} \\
z_{5,3} \\
f:26 \\
\bot
\end{gather*}}}
]
]
[.\ensuremath{\parbox{.08\columnwidth}{\begin{gather*}
5:~n_7 \\
z_{5,1} \\
f:13 \\
\bot
\end{gather*}}}
]
[.\ensuremath{\parbox{.08\columnwidth}{\begin{gather*}
5:~n_8 \\
z_{5,3} \\
f:26
\end{gather*}}}
]
[.\ensuremath{\parbox{.08\columnwidth}{\begin{gather*}
5:~n_9 \\
z_{5,4} \\
f:16 \\
\bot
\end{gather*}}}
]
]
[.\ensuremath{\parbox{.08\columnwidth}{\begin{gather*}
4:~n_5 \\
z_{5,3} \\
f:26
\end{gather*}}}
]
]
[.\ensuremath{\parbox{.18\columnwidth}{\begin{gather*}
2:~n_2  \\
z_{3,3}\\
f:10 \\
\bot
\end{gather*}}}
]
[.\ensuremath{\parbox{.18\columnwidth}{\begin{gather*}
2:~n_3 \\
z_{4,4} \\
f:33 
\end{gather*}}}
[.\ensuremath{\parbox{.18\columnwidth}{\begin{gather*}
3:~n_3  \\
\emptyset\\
f:12 \\
\bot
\end{gather*}}}
]
]
]
\caption{Search tree for Example~\ref{ex:last-nested-sections}}
\label{fig:search-tree}
\end{footnotesize}
\end{figure}

%% file: conclusion.tex
We have introduced a polynomial method for bounding the blocking time, and exact method for computing the blocking time under PIP. There is surely margin for further optimizations. For example, dynamic programming techniques can be used to cache partial results on the admissibility of z-chains. Nevertheless, the approach we propose already offers two major benefits: it shows that establishing a bound can be done in polynomial time, whereas literature  has only offered, to the best of our knowledge, exponentially-complex methods, and it defines an exact method, which was something missing altogether. Moreover, the proposed method is optimally efficient. A further contribution is the first complete characterization of blocking under PIP, which could lay the ground for further analyses and a better understanding of the theory and practice of such a key component of many real-time systems.